\documentclass[%
 reprint,
 amsmath,amssymb,
 aps,
floatfix,
]{revtex4-2}

\usepackage{graphicx}
\usepackage{dcolumn}
\usepackage{bm}
\usepackage{color}
\usepackage{float}
\usepackage{upgreek}
\usepackage{CJK}
\usepackage[%
  colorlinks=true,
  urlcolor=blue,
  linkcolor=blue,
  citecolor=blue
]{hyperref}
\usepackage{siunitx}

\DeclareSIUnit\electrons{e\textsuperscript{--}}
\DeclareSIUnit\neutrons{neutrons}
\DeclareSIUnit\ppm{ppm}
\DeclareSIUnit\ppb{ppb}
\DeclareSIUnit\lines{l}
\DeclareSIUnit{\calorie}{cal}

\begin{document}

\begin{CJK*}{UTF8}{gbsn}

\preprint{APS/123-QED}

\title{Phosphorescence and donor-acceptor pair recombination in laboratory-grown diamonds}

\author{Jiahui Zhao (赵嘉慧)}
 \email{gloria.zhao.1@warwick.ac.uk}
 \affiliation{Department of Physics, University of Warwick, Coventry CV4 7AL, United Kingdom
}
\author{Ben G. Breeze}
  \affiliation{Spectroscopy Research Technology Platform, University of Warwick, Coventry CV4 7AL, United Kingdom}
\author{Ben L. Green}
\affiliation{Department of Physics, University of Warwick, Coventry CV4 7AL, United Kingdom
}
\author{Mark E. Newton}
 \email{m.e.newton@warwick.ac.uk}
\affiliation{Department of Physics, University of Warwick, Coventry CV4 7AL, United Kingdom
}
\maketitle
\end{CJK*}




\date{\today}

\begin{abstract}
 Intense ``blue-green" phosphorescence is commonly observed in near colourless lab-grown high-pressure high-temperature (HPHT) diamonds following optical excitation at or above the indirect bandgap. We have employed a holistic combination of optically-excited time-resolved techniques (in addition to standard spectroscopic characterisation techniques) to study the physics of this long-lived phosphorescence and understand luminescence-related charge transfer processes.

It is shown that the properties of the broad ``blue-green" luminescence and phosphorescence band can be fully explained by emission from neutral substitutional nitrogen-boron donor-acceptor pairs ($\text{N}_\text{S}^0$...$\text{B}_\text{S}^0$) , once the configurational change between charge states is considered, and both tunneling between defects and thermal ionization of donors and acceptors is considered. Significant concentrations of metastable $\text{N}_\text{S}^-$, are identified after optical excitation at or above the indirect bandgap. $\text{N}_\text{S}^-$ is much shallower ($\sim$ 0.2~eV) than previously thought and plays a key role in resetting the $\text{N}_\text{S}^0$...$\text{B}_\text{S}^0$ donor-acceptor pairs.   
\end{abstract}

\maketitle


\section{Introduction}

Diamond is a wide-band-gap material (E$_\text{G}$ = 5.47 eV; indirect) with large donor (Nitrogen: E$_\text{D}$ = 1.7 eV{\cite{farrer1969substitutional}}; Phosphorus: E$_\text{D}$ = 0.62 eV\cite{Stenger2013}) and acceptor (Boron: E$_\text{A}$ = 0.368 eV\cite{collins_71}) ionization energies. If, during the production of diamond by High Pressure High Temperature (HPHT) synthesis, a sufficient concentration of nitrogen getters\cite{Sonin2022} is added to the growth capsule the concentration of substitutional nitrogen ($\text{N}_\text{S}$) impurities incorporated into the diamond can be reduced to less than \SI{200}{\ppb} (parts per billion carbon atoms). The resulting diamond samples typically contain similar concentrations of substitutional boron ($\text{B}_\text{S}$) impurities. Such diamonds would be classified as type II {\footnote{Type II: diamond contains very low level of nitrogen impurities ($<$ 1 ppm)}}. Phosphorescence, or delayed luminescence, is commonly observed in such HPHT diamonds after illumination with light of energy equal to or greater than the indirect bandgap. The phosphorescence lasts for a few milliseconds up to several hours after the removal of optical excitation {\cite{watanabe1997phosphorescence,eaton2016}}. Combinations of phosphorescence imaging and other spectroscopic identification techniques are routinely used to distinguish lab-grown HPHT diamonds from natural diamonds {\cite{Ulrika2015large}}. 

A ``blue-green" phosphorescence band, with a peak energy of 2.5(1) eV (\SIrange{490}{503}{\nm}) is often observed in type II natural, HPHT synthetic, and HPHT treated diamonds {\cite{watanabe1997phosphorescence,eaton2016,eaton2011phosphorescence,eaton2008using,Ulrika2015large,walsh1971thermoluminescence,song2016identification,chandrasekharan1946patterns,chandrasekharan1946TL,krumme1964}}. There is a lack of research on phosphorescence in CVD synthetic diamonds. Almost all phosphorescence in natural, HPHT synthetic and HPHT treated diamonds detected after bandgap excitation at room temperature contains the ``blue-green" component. The ``blue-green" phosphorescence is quenched at higher temperatures, and the band position is temperature independent in a range of \SIrange{250}{400}{\kelvin} {\cite{krumme1964,su2018}}. The ``blue-green" emission band was not seen after excitation with light of wavelengths 365 nm and 458 nm {\cite{watanabe1997phosphorescence,eaton2008using}}. 

In previous work, the ``blue-green" phosphorescence has been suggested to originate from temperature-dependent ionisation of substitutional boron acceptors combined with recombination emission at donor-acceptor pairs (DAPs) {\cite{watanabe1997phosphorescence}}. However, questions about the defects involved and the details of the mechanism remain. The incorporation of defects/impurities in HPHT diamond is growth-sector dependent, and the differences in phosphorescence behaviour between growth sectors significantly complicates analysis in multi-sectored samples. The nature of the donor and the involvement of other defects or luminescence centres needs to be investigated. The temperature-dependence of phosphorescence has not been fully investigated: phosphorescence observed at liquid nitrogen temperature cannot be explained by the thermal excitation mechanism proposed in the literature (see \S{\ref{sec:phos}} below). Furthermore, it is somewhat surprising that low concentrations of impurities/defects in type II HPHT diamonds are sufficient to give rise to the intense phosphorescence that is often observed.

The energy obtained from electromagnetic or ionizing radiation (such as optical excitation, x-rays or electron irradiation) can be stored in the diamond by trapping charge carriers at defects. After excitation, a fraction of the stored energy can be released as phosphorescence; in other cases, the energy can be stored in the dark for a long time and is released as thermoluminescence (TL) when heated, via thermally-activated de-trapping of charge carriers {\cite{chandrasekharan1946patterns,mckeever1988thermoluminescence}}. Therefore, TL is a useful technique to investigate the interaction between traps and optically active defects in diamond. The shape, position, and intensity of the TL emission (``glow curve") indicates the depth and the relative concentration of traps {\cite{mckeever1988thermoluminescence,paslovsky1993interpretation}}. ``Blue-green" and ``red" TL is seen in natural, HPHT synthetic, and boron doped CVD synthetic diamonds, regardless of type I or type II {\cite{chandrasekharan1946patterns,petitfils2007role,walsh1971thermoluminescence,halperin1966thermoluminescence,levinson1973electrically,bourgoin1978thermally,bull1950luminescence,halperin1961some}}. The ``blue-green" TL band centred at  \SI{2.57}{\eV} {\cite{petitfils2007role}} or \SI{2.6}{\eV} {\cite{walsh1971thermoluminescence}} seen after UV illumination could be linked to the ``blue-green" phosphorescence band observed by {Shao \emph{et al.}} {\cite{shao2020role}}. TL emission which peaks at \SI{480}{\kelvin} after UV illumination in a weakly boron-doped type IIb CVD synthetic diamond film has a broad spectral band centred at $\sim$ 490 nm, close to the ``blue-green" phosphorescence band position {\cite{benabdesselam2000characterisation}}. Thus the same defects may play roles as luminescence centres in both phosphorescence and thermoluminescence {\cite{walsh1971thermoluminescence}}. Two TL glow peaks corresponding to traps with activation energies of approximately \SI{0.2}{\eV} and \SI{0.37}{\eV} are commonly observed in type II natural and HPHT synthetic diamonds {\cite{walsh1971thermoluminescence,nahum1963thermoluminescence,halperin1966thermoluminescence,levinson1973electrically,bourgoin1978thermally}}. The TL glow peak of activation energy of \SI{0.37}{\eV} is widely agreed to be assigned to substitutional boron acceptor. However, the trap identified from lower activation energy TL peak is uncertain: Walsh considered it as a shallower acceptor {\cite{walsh1971thermoluminescence}}, while Bourgoin believed its energy level is approximately \SI{0.22}{\eV} below the conduction band {\cite{bourgoin1978thermally}}. It is notable that this TL glow peak is more intense in diamond samples containing a higher concentration of nitrogen {\cite{walsh1971thermoluminescence}}.

In this research, time-resolved Fourier-transform infrared (FTIR) absorption and time-resolved electron paramagnetic resonance (EPR) techniques are used to understand the role of $\text{B}_\text{S}$  and $\text{N}_\text{S}$ defects in the charge transfer processes, in order to explain the phosphorescence and thermoluminescence behaviour in the HPHT lab-grown diamond samples studied. 
The incorporation efficiency of boron is highly growth-sector dependent in HPHT lab-grown diamonds {\cite{howell2019automated,blank2007influence,davies1994properties}}. Higher HPHT crystallization temperature results in increased uncompensated boron concentration in the \{111\} and \{100\} growth sector {\cite{blank2007influence}}. The rate of uptake boron by various growth sectors of HPHT diamond is \{111\} $>$ \{110\} $>$ \{100\}, \{113\} {\cite{burns1990growth}}. In HPHT diamonds, the preference for nitrogen incorporation is also growth sector dependent, and the nitrogen content usually follows \{111\} $>$ \{100\} $>$ \{113\} $>$ \{110\} {\cite{ashfold2020nitrogen,burns1990growth}}. Substitutional nitrogen $\text{N}_\text{S}$ is also known as the C-centre or P1 defect in EPR studies. 

In principle, at room temperature and below, the Fermi energy of an N- or B-doped diamond is pinned close to the $\text{N}_\text{S}^0$ donor or $\text{B}_\text{S}^0$ acceptor level, depending on which is present in the higher concentration. Due to the relatively deep donor / acceptor levels, the probability of thermally ionizing the $\text{N}_\text{S}^0$ donor is negligible, and only a very small faction ($<1\%$) of the $\text{B}_\text{S}^0$ acceptors will be ionized at room temperature. However, in an insulating material like diamond the calculated position of the Fermi level does not necessarily predict the correct charge state of a defect, and multiple charge states of the same defect can be present at the same time {\cite{collins2002fermi}}. In this situation the charge state of a defect is influenced by its proximity to a donor (or acceptor) and the optical excitation / thermal history of the diamond.  Thus, $\text{N}_\text{S}^0$, $\text{N}_\text{S}^+$, $\text{B}_\text{S}^0$ and $\text{B}_\text{S}^-$ are are all expected to be present in a diamond containing low concentrations of both $\text{B}_\text{S}$ and $\text{N}_\text{S}$. Furthermore, substitutional nitrogen has been predicted to have an acceptor level ($\text{N}_\text{S}^-$) lying approximately 1.1 eV below the minimum of the conduction band.{\cite{jones2009acceptor}} Experimental evidence for $\text{N}_\text{S}^-$ was provided by ultra-fast spectroscopic measurements in which a transient infrared absorption feature at 1349 cm$^{-1}$ was assigned to a local vibrational mode of $\text{N}_\text{S}^-$ {\cite{ulbricht2011single}}. 

In previous studies of charge transfer and phosphorescence in near colourless HPHT grown diamond, the role of $\text{N}_\text{S}^-$ has not been considered, while in this work the existence of $\text{N}_\text{S}^-$ is shown to be crucial for the interpretation of the data and is found to be a much shallower donor than previously predicted {\cite{jones2009acceptor}}.

\section{Experimental details}

\subsection{Samples}

The single-crystal diamond samples GE81-107a-B and GE81-107a-C were grown using the temperature gradient method by General Electric with iron cobalt solvent-catalyst and an aluminium nitrogen getter. From the as-grown sample a \{110\} oriented plate ($\sim$ 0.85 mm thick) was prepared and subsequently cut and polished into separate samples: GE81-107a-B (B\{001\} henceforth) is dominated by \{100\} growth sector but also contains some other minor growth sectors, and GE81-107a-C (C\{111\} henceforth) consists of a \{111\} growth sector. Both samples were near colourless but weakly boron-doped and present characteristic ``blue-green" fluorescence under bandgap UV excitation, with subsequent long-lived phosphorescence. The average concentration of $\text{B}_\text{S}^0$ and $\text{N}_\text{S}^0$ in both samples in the ``meta-stable ambient state" (e.g. at room temperature after daylight/laboratory illumination for $>$ 1 hour) were determined by FTIR absorption and EPR spectroscopy (see TABLE {\ref{Table: GE samples concentrations}}).

\begin{table}[H]
\renewcommand\arraystretch{1.3}
\centering
\caption{The concentrations of $\text{B}_\text{S}^0$ and $\text{N}_\text{S}^0$ in samples B\{001\} and C\{111\} in the ``meta-stable ambient state".}
\begin{ruledtabular}
\begin{tabular}{ccc}
Concentration (ppb) & B\{001\} & C\{111\} \\ \hline
$[\text{B}_\text{S}^0]$ by FTIR & 67 $\pm$ 10 & 332 $\pm$ 40 \\
$[\text{N}_\text{S}^0]$ by EPR & 145 $\pm$ 20 & $<$ 4 \\ 
\end{tabular}
\label{Table: GE samples concentrations}
\end{ruledtabular}
\end{table}

\subsection{Time-resolved FTIR}

To study $\text{B}_\text{S}$-related charge transfer during phosphorescence decay, an optically-excited variable temperature FTIR absorption experiment was set up using a Thermo Fisher Scientific Nicolet iS50R FTIR spectrometer. The diamond sample was cooled or heated to a selected temperature between 273 K to 573 K by a variable temperature stage (Linkam THMS600), and a reference spectrum was collected with the sample in the dark. A \SI{224}{\nano\meter} pulsed laser (Photon Systems HEAG70-224SL) operating at 10~Hz was focused on the sample until the  $\text{B}_\text{S}^0$ absorption reached a steady state under optical excitation (saturation). After saturation was achieved the optical excitation was removed, and the FTIR spectrometer triggered to collect spectra every 210 ms by an Arduino.  Thus the decay of the \SI{2802}{\per\centi\meter} $\text{B}_\text{S}^0$ absorption peak was recorded during the phosphorescence decay \cite{collins2010}.

\subsection{Time-resolved EPR}

The decay/recovery of the $\text{N}_\text{S}^0$ concentration post-optical-excitation was monitored in an optically-excited variable temperature EPR experiment utilising a Bruker E580 spectrometer equipped with an X-band microwave bridge. A Bruker ER4131 VT temperature controller was used to cool or heat the sample to a temperature between \SIrange{160}{400}{\kelvin}. After optical excitation to saturation (i.e., no further change in $\text{N}_\text{S}^0$ EPR signal) with the \SI{224}{\nano\meter} laser, the EPR spectrometer was triggered to start rapid-passage scans in the dark at a sweep frequency of 10~Hz and sweep width of 2~mT across the central line of the $\text{N}_\text{S}^0$ EPR spectrum. A National Instruments CompactRIO controller combined with NI-9219 and NI-9264 modules was used to control the drive of the field sweep coils (sweep rate up to 200~mT/s), and ensure synchronisation with the EPR data acquisition {\cite{breeze2016electron}}.

\subsection{Luminescence}

An Edinburgh Instruments FS5 spectrofluorometer was used to measure the excitation dependence of photoluminescence in the diamonds at room temperature over an excitation range of \SIrange{200}{500}{\nm} and an emission range of \SIrange{200}{800}{\nm}.

Cathodoluminescence spectra were recorded by a scanning electron microscope equipped with a Gatan Mono CL system at 80 K and room temperature.

\begin{figure}[!ht]
    \centering
    \includegraphics[scale=0.35]{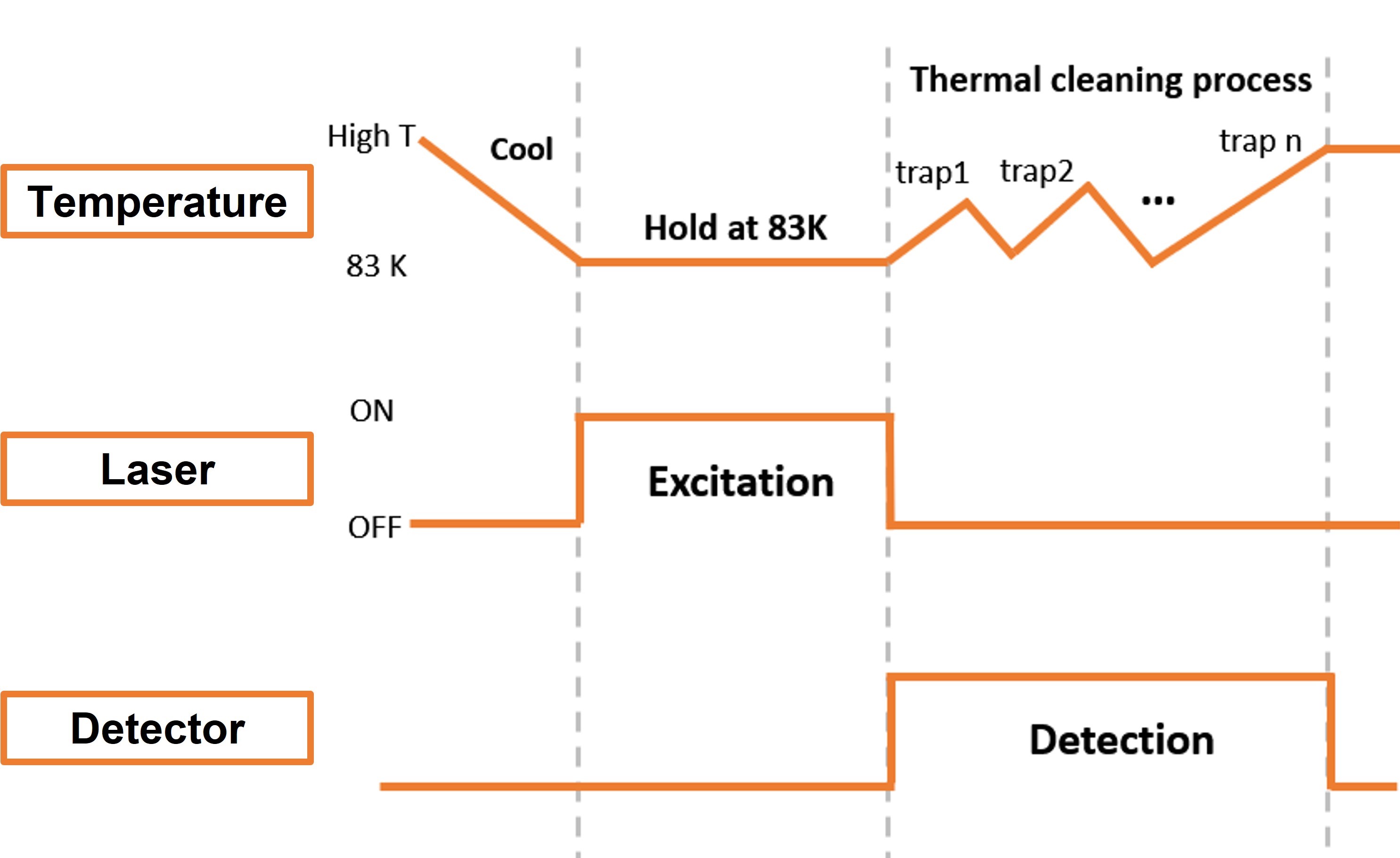}
    \caption{Schematic of the thermoluminescence (TL) experiment sequence for deconvolving multiple overlapping TL peaks (known as ``thermal cleaning" \cite{mckeever1988thermoluminescence}).}
    \label{fig:TL experiment}
\end{figure}

An experimental system was built to study phosphorescence and TL in diamond, which enabled time-resolved bulk spectral and hyperspectral acquisition, when combined with the \SI{224}{\nano\meter} pulsed laser. The sample was mounted in a variable temperature stage (Linkam THMS600) with the working temperature range between 83 K and 873 K. Sample emission was detected by either a fibre-coupled spectrometer (ANDOR Shamrock i303) operating between \SIrange{400}{1000}{\nm}, or a camera (HAMAMATSU CMOS C11440-36U) sensitive to \SIrange{300}{1100}{\nm}. The hardware components, including the laser, temperature stage, filters and detectors, were controlled by an Arduino Uno, which enabled $\sim$ms time accuracy in phosphorescence and thermoluminescence experiments. To perform a phosphorescence experiment, the diamond sample was cooled or heated to a selected temperature in absence of optical excitation and then optically excited to saturation of the initial phosphorescence emission before detection was triggered. When performing a TL experiment, the sample was cooled to 83 K and optically excited to saturation. After the light source was removed, the sample was then heated up to 473 K at a linear rate of 100 K/min after a variable delay time. TL glow was recorded by the camera or spectrometer. The ``thermal cleaning" method (Fig {\ref{fig:TL experiment}}) was used to separate overlapping TL glow peaks: after initial optical excitation at \SI{83}{\kelvin}, the sample was heated as in a normal TL experiment, but cooled back down to \SI{83}{\kelvin} just after the maximum intensity of each distinct TL peak was reached. This procedure was repeated until all peaks were detected {\cite{mckeever1988thermoluminescence}}.

\section{Results}

\subsection{\texorpdfstring{$\text{B}_\text{S}^0$}{Bs0} related charge transfer}

\begin{figure}[!ht]
    \centering
    \includegraphics[scale=0.45]{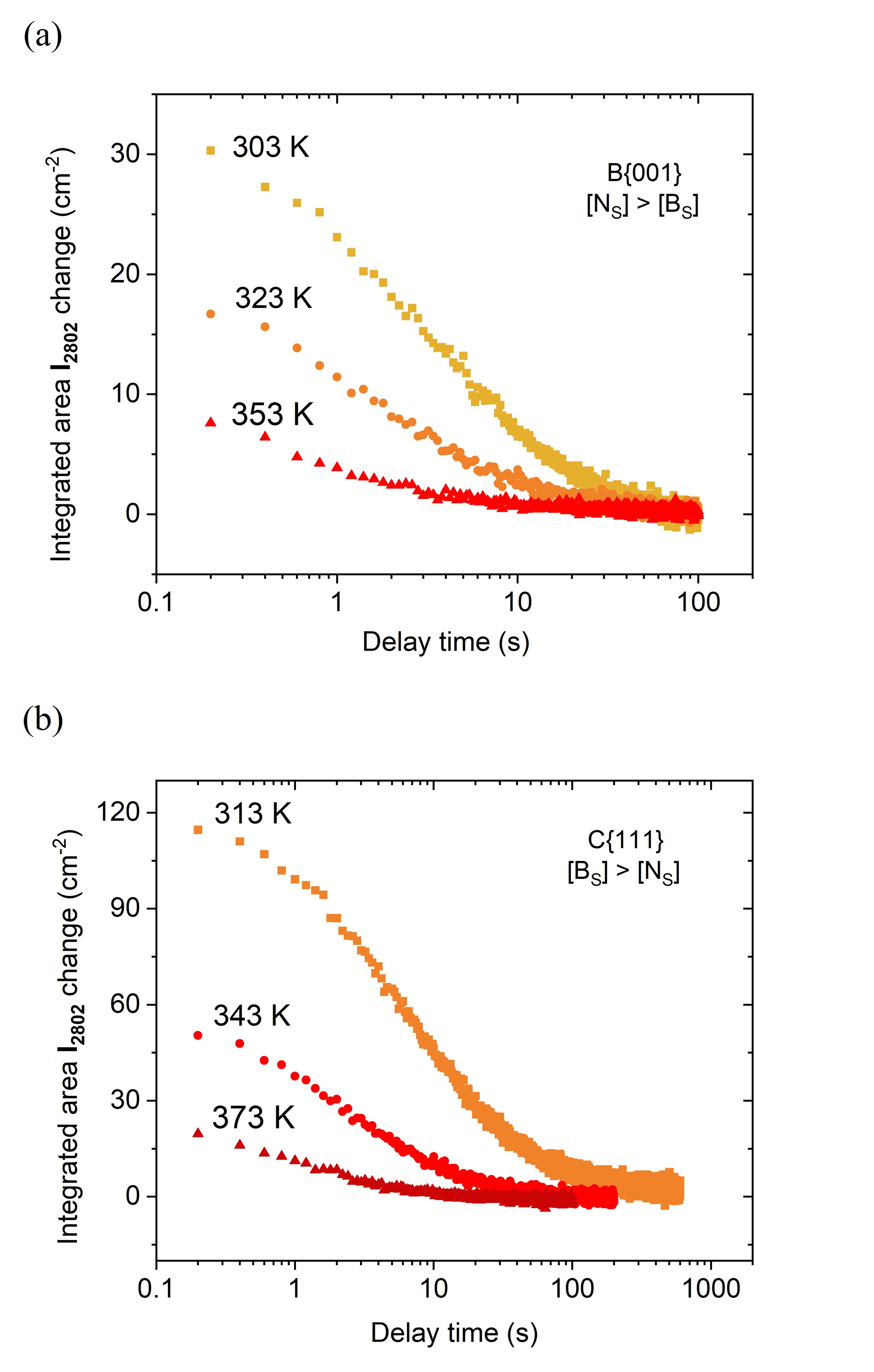}
    \caption{The integrated area of 2802 $\text{cm}^{-1}$ peak recorded post-illumination with a \SI{224}{\nano\meter} laser for (a) sample B\{001\} where [$\text{N}_\text{S}$] $>$ [$\text{B}_\text{S}$], and (b) C\{111\} where [$\text{B}_\text{S}$] $>$ [$\text{N}_\text{S}$]. In both samples, the $\text{B}_\text{S}$ concentration decreases once optical excitation is removed.}
    \label{fig:GE FTIR decay curves}
\end{figure}

Above bandgap UV excitation increased the concentration of uncompensated substitutional
boron in both samples, as measured by FTIR. As the sample temperature was increased, the UV illumination had less effect on the $\text{B}_\text{S}^0$ concentration. Post-excitation decay curves of the \SI{2802}{\per\centi\meter} $\text{B}_\text{S}^0$ absorption peak intensity show the decay rate is temperature-dependent [Fig {\ref{fig:GE FTIR decay curves}}]. The data is plotted on a logarithm time scale to emphasise the change in the decay rates. Consistent with literature reports, the $\text{B}_\text{S}^0$ 2802 cm$^{-1}$ absorption peak decays during phosphorescence {\cite{li2016diamond,eaton2016decay}}. At each temperature, the $[\text{B}_\text{S}^0]$ decay rate in the \{111\} growth sector is slower than that in the (lower boron concentration) \{001\} growth sector.

\subsection{\texorpdfstring{$\text{N}_\text{S}^0$}{Ns0} related charge transfer}

\begin{figure}[!ht]
    \centering
    \includegraphics[scale=0.45]{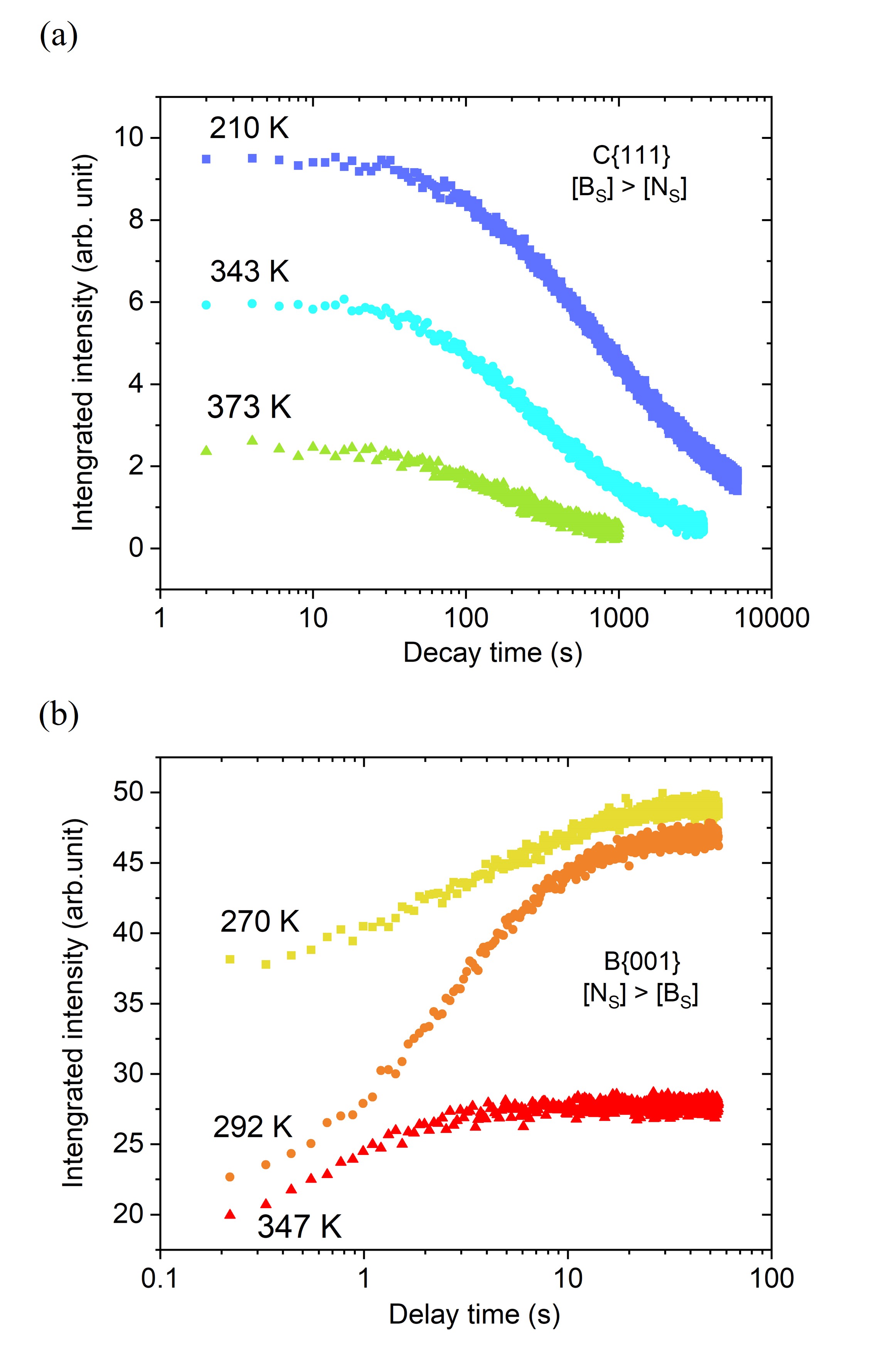}
    \caption{Time-resolved integrated intensity of the central ${\text{N}_\text{S}^0}$ hyperfine peak following \SI{224}{\nano\meter} illumination, as measured by rapid-passage EPR for sample (a) C\{111\} where [$\text{B}_\text{S}$] $>$ [$\text{N}_\text{S}$]; and (b) B\{001\} where [$\text{N}_\text{S}$] $>$ [$\text{B}_\text{S}$].}
    \label{fig:GE EPR decay curves}
\end{figure}

Rapid-passage, post-UV-illumination measurements of the central peak of $\text{N}_\text{S}^0$ EPR spectrum (plotted as an integrated intensity) in sample C\{111\} [Fig {\ref{fig:GE EPR decay curves}(a)}] show that [$\text{N}_\text{S}^0$] decreases post-illumination in this sample. The lifetime of the $\text{N}_\text{S}^0$ concentration decay is temperature dependent: the higher the temperature, the shorter the lifetime (the lifetime varies by a factor of $>5$ between 210 K to 250 K). 

Conversely, the metastable $\text{N}_\text{S}^0$ concentration in sample B\{001\} is reduced by 224 nm laser excitation in the temperature range \SIrange{270}{347}{\kelvin}: the central peak of the $\text{N}_\text{S}^0$ EPR spectrum recovers on a time scale of tens of seconds after the removal of UV illumination [Fig {\ref{fig:GE EPR decay curves}(b)}]. The recovery rate is strongly temperature-dependent between \SI{270}{\kelvin} and \SI{318}{\kelvin}, whereas it only becomes slightly faster as the temperature is increased above 318 K. It should be noted that above 290 K the metastable $\text{N}_\text{S}^0$ concentration is strongly temperature dependent, decreasing as the temperature is increased.

\subsection{Luminescence}

\subsubsection{Photoluminescence}

\begin{figure}[!ht]
    \centering
    \includegraphics[scale=0.45]{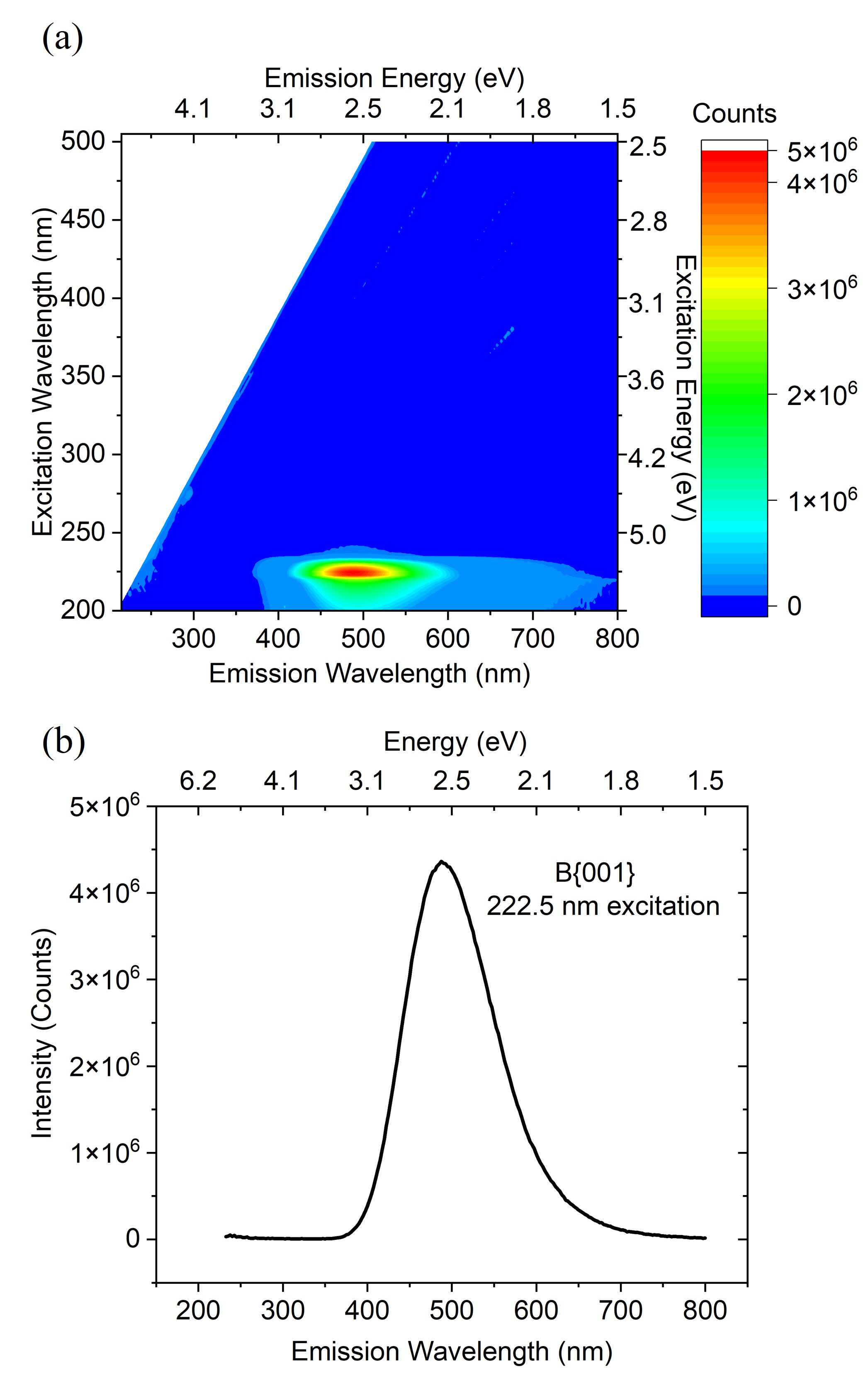}
    \caption{(a) Photoluminescence excitation spectra of sample B\{001\} at room temperature under excitation ranging from 200 to \SI{500}{\nano\meter}. (b) Photoluminescence spectrum under \SI{222.5}{\nano\meter} excitation.}
    \label{fig:GE PL}
\end{figure}

Two-dimensional room temperature photoluminescence excitation experiments allow us to measure the excitation dependence of the ``blue-green" luminescence band centred at approximately 2.5 eV in the diamond samples studied [Fig {\ref{fig:GE PL}(a)}]. The centre of the band shifts from \SI{490}{\nano\meter} to \SI{480}{\nano\meter} as the excitation wavelength varies from \SIrange{200}{235}{\nano\meter}. The luminescence band is most intense under $\sim$ 222.5 nm excitation [Fig {\ref{fig:GE PL}(b)}]. It is clear that in order to excite the ``blue-green" emission, an excitation wavelength shorter than $\sim$ 235 nm is required. Interestingly this corresponds to the free exciton emission energy in a diamond. Thus it appears that excitation to produce free electrons and holes is required to observe the ``blue-green" emission: radiative relaxation can then occur via reasonably close donor-acceptor pair recombination.

\subsubsection{Cathodoluminescence}

\begin{figure}[!ht]
    \centering
    \includegraphics[scale=0.30]{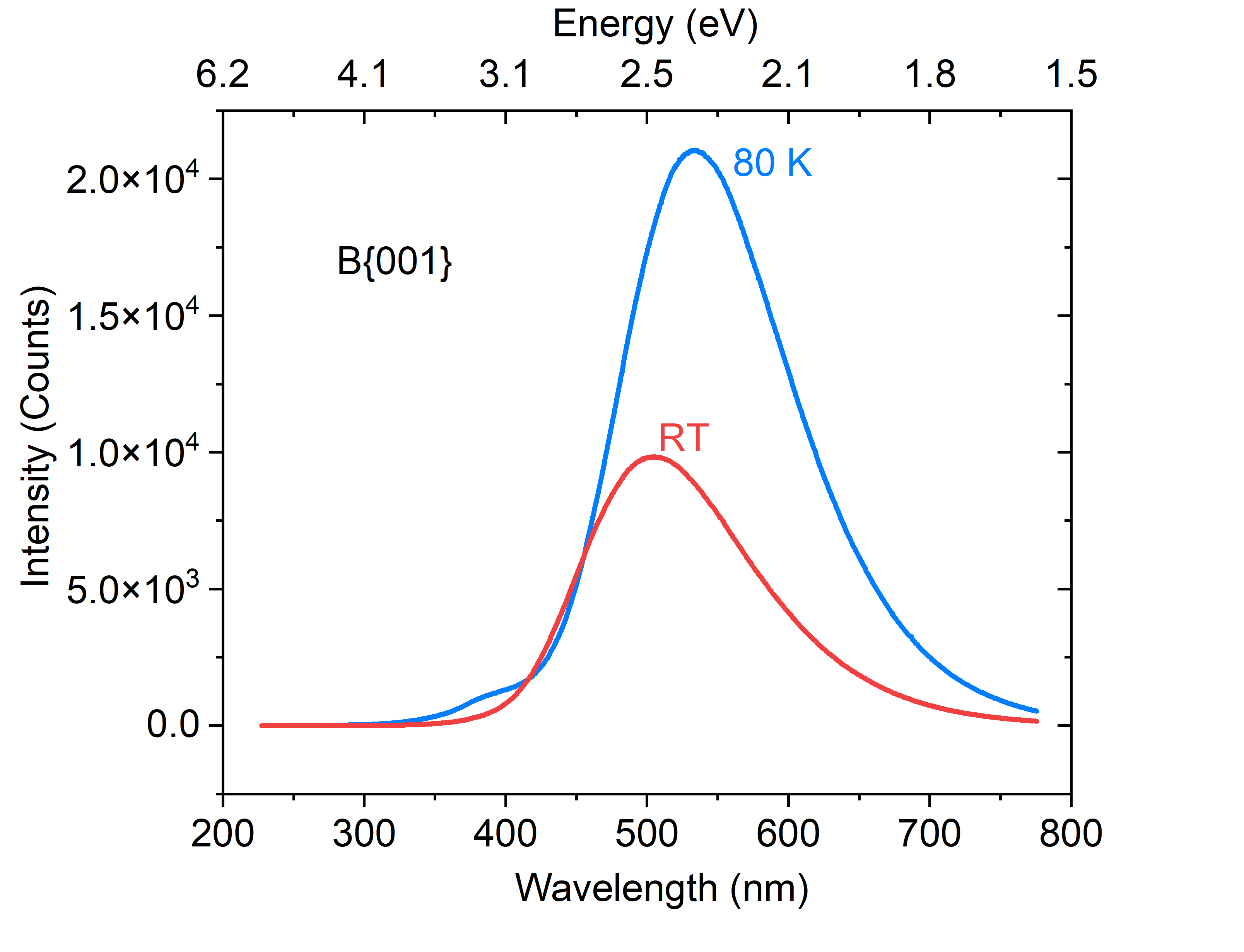}
    \caption{CL spectra of sample B\{001\} at 80 K and room temperature.}
    \label{fig:GE CL}
\end{figure}

Cathodoluminescence spectra from both growth sectors at 80 K and room temperature shows strong emission bands, and the peak intensity shifts to lower energy as the sample temperature is reduced. The CL emission band from sample B\{001\} is centred at 532 nm (\SI{2.33}{\eV}) and \SI{503}{\nano\meter} (\SI{2.46}{\eV}) at \SI{80}{\kelvin} and room temperature, respectively [Fig {\ref{fig:GE CL}}]; the CL emission band in sample C\{111\} is peaked at 535 nm (2.32 eV) and 520 nm (2.28 eV) at \SI{80}{\kelvin} and room temperature, respectively.

\subsubsection{Phosphorescence}                         \label{sec:phos}

\begin{figure}[!ht]
    \centering
    \includegraphics[scale=0.45]{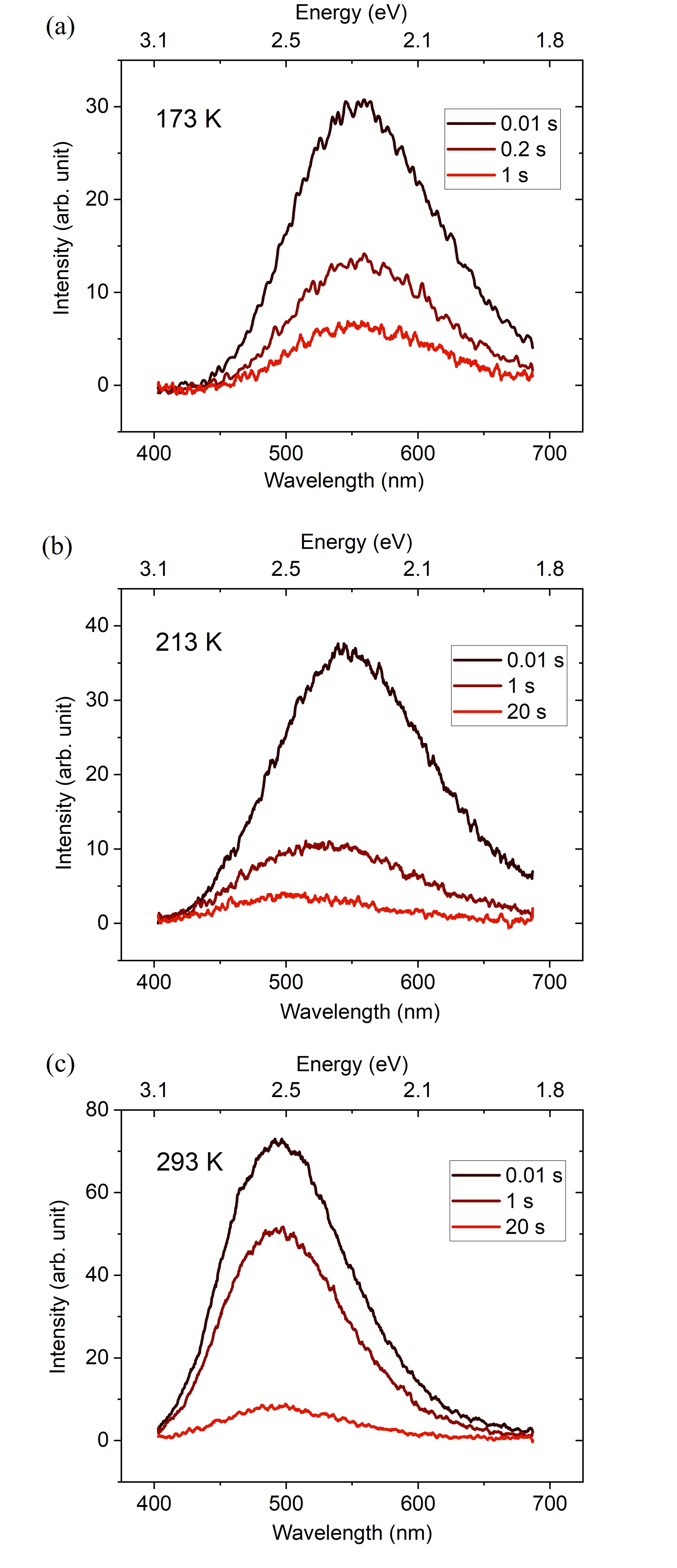}
    \caption{Phosphorescence spectra (labelled by the delay time between switching off the laser and starting acquisition) of sample C\{111\} recorded post-224~nm excitation at (a) 173 K, (b) 233 K, and (c) 293 K. }
    \label{fig:GE Phos spectra}
\end{figure}

Characteristic ``blue-green" phosphorescence is observed in both samples post-224 nm excitation between 83 K and 473 K. The phosphorescence band in sample C\{111\} peaks at 2.25~eV (550 nm) in the low-temperature range of \SIrange{83}{173}{\kelvin}, and approximately 2.5 eV (494 nm) at high temperature between 273 K and 473 K [Fig~{\ref{fig:GE Phos spectra}}]. There is no shift in the band peak energies during decay in either of these low (\SIrange{83}{173}{\kelvin}) or high (\SIrange{273}{473}{\kelvin}) temperature ranges. The 2.5 eV band corresponds to the phosphorescence reported in the literature {\cite{eaton2011phosphorescence,watanabe1997phosphorescence,eaton2008using,Ulrika2015large,walsh1971thermoluminescence}}. In the intermediate temperature range (\SIrange{173}{273}{\kelvin}), the emission band position shifts to higher energy significantly during decay, which suggests the mechanisms of phosphorescence switch within this temperature range and the emission consists of at least two components: a phosphorescence band peaked at lower energy which decays faster than a phosphorescence band peaked at higher energy. 

The temperature dependence of the emission band position in both samples is similar. At low temperatures (\SIrange{83}{173}{\kelvin}), the phosphorescence band is also centred at 2.25 eV (550 nm) in B\{001\}, same as that in C\{111\}. At high temperatures above 273 K, the ``blue-green" phosphorescence band in B\{001\} is centred at 484 nm, slightly shorter in wavelength than that of C\{111\} (494 nm).

\begin{figure*}[t]
    \centering
    \includegraphics[scale=0.50]{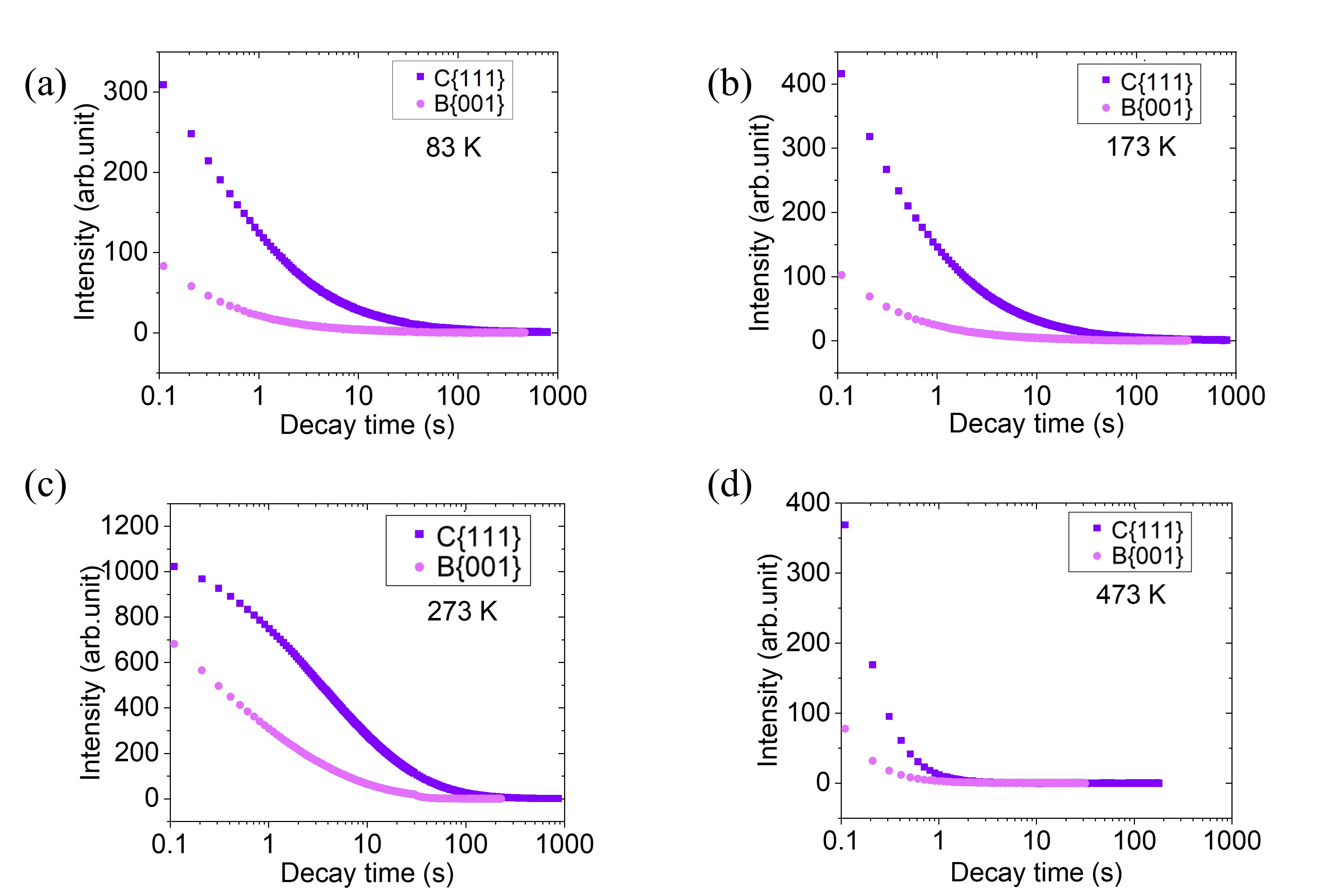}
    \caption{Phosphorescence intensity decay curves recorded (50 ms acquisition time for each frame) post-224 nm-excitation for samples C\{111\} and B\{001\} at different temperatures.}
    \label{fig:GE phos decay different T}
\end{figure*}

In our experiment, phosphorescence intensities are plotted in arbitrary units, but nevertheless a comparison can be made between the relative intensities in both samples since the acquisition parameters are the same for all measurements [Fig {\ref{fig:GE phos decay different T}}]. The phosphorescence was more intense in C\{111\} than that in B\{001\} at all temperatures studied, but the difference was largest for the measurements in the low and high temperature ranges. The phosphorescence decay rate and emission spectrum at low temperatures (\SIrange{83}{173}{\kelvin}) in both samples are approximately temperature-independent. At temperatures above \SI{173}{\kelvin}, the phosphorescence lifetime decreased rapidly as the temperature was increased until the decay was too fast to be detected by the experimental setup above \SI{473}{\kelvin}.

A multiple component fitting method (see {\ref{apx:multiple}}) was used to interpret the dominant phosphorescence mechanism at different temperatures. The phosphorescence decay curves at low temperatures (\SIrange{83}{173}{\kelvin}) can be satisfactorily fit using two athermal power-law decay components with different parameters, which suggests that the phosphorescence decay in this temperature range originates from athermal tunnelling processes. At temperatures above \SI{273}{\kelvin}, the decay can be fitted using two simple exponential curves, suggesting the phosphorescence at high temperature is dominated by thermal activation. Notably, better fitting is achieved by simple exponential functions than hyperbolic functions, which suggests the thermal processes follow first-order kinetics and charge retrapping is negligible. At an intermediate temperature (\SIrange{173}{273}{\kelvin}), the phosphorescence decay curve is best fitted using a combination of athermal power-law decay functions and simple exponential functions, suggesting that both tunnelling and thermal activation contributes to phosphorescence at these temperatures.

\subsubsection{Thermoluminescence}

\begin{figure}[!ht]
    \centering
    \includegraphics[scale=0.30]{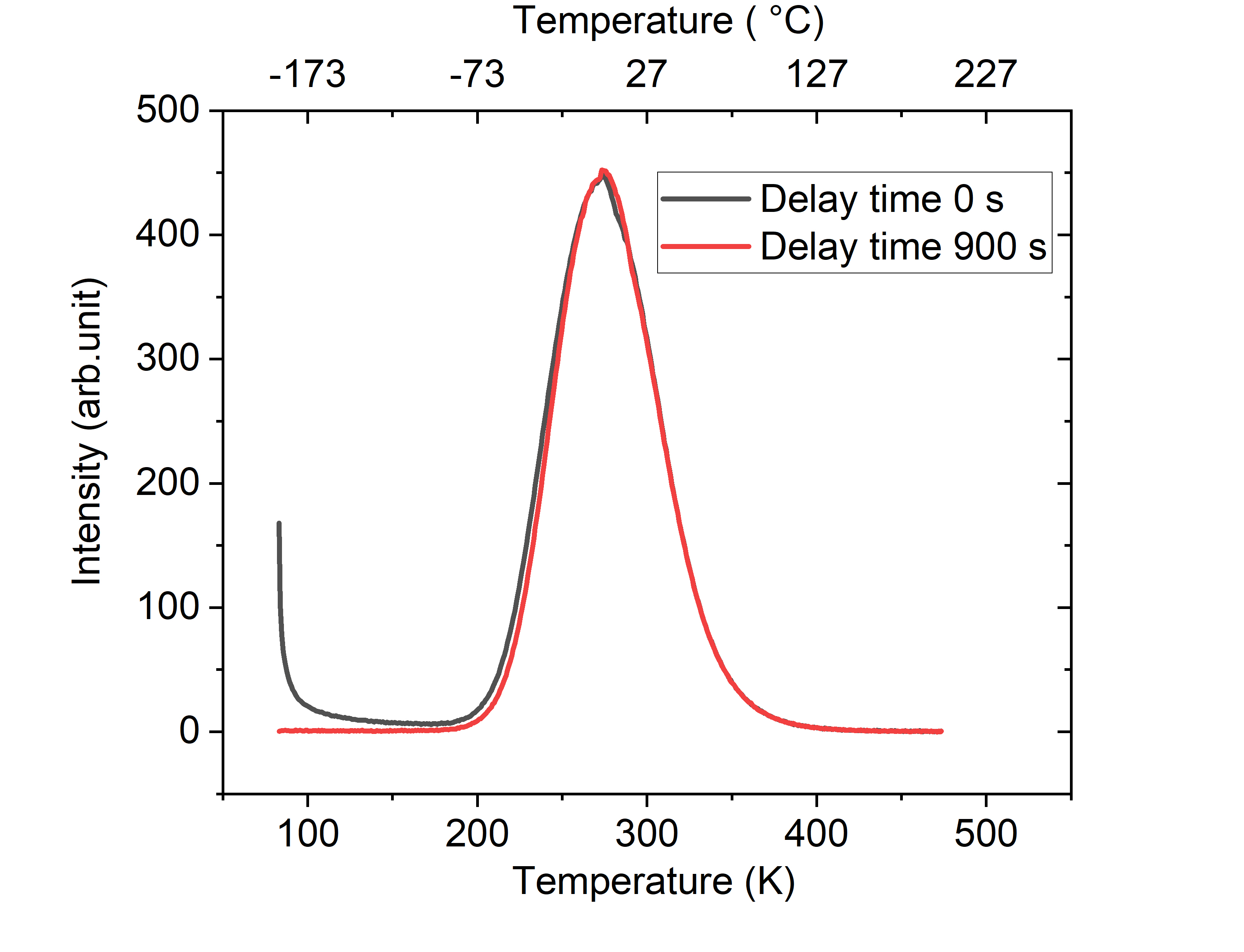}
    \caption{By delaying the start of the TL temperature ramp, defect centres which relax via phosphorescence can be eliminated from the subsequent TL glow curve. Here, sample C\{111\} was excited using a 224~nm laser at \SI{83}{\kelvin} and then heated at \SI{100}{\kelvin\per\minute} after the indicated post-laser delay.}
    \label{fig:GE TL with and without decay}
\end{figure}

Even at low temperature, following \SI{224}{\nano\meter} excitation there is significant phosphorescence generated: this can interfere with thermoluminescence measurements if the temperature is ramped immediately post-illumination. By delaying the start of the temperature ramp, we ensure the phosphorescence has completely decayed before beginning the thermoluminescence measurement [Fig {\ref{fig:GE TL with and without decay}}]. As a result, the detected thermoluminescence only arises from the luminescence centres which are located far from the traps and luminescence that arises from the tunnelling process (i.e., near pairs which relax via phosphorescence) was excluded. 

The TL intensity in C\{111\} is stronger than that in B\{001\}, while the shape and position of the TL glow curves are similar: a single asymmetric TL glow peak was observed from 173 K to approximately 400 K and centred at 273 K. The low temperature side of the TL peak (\SIrange{173}{273}{\kelvin}) corresponds to the ``intermediate temperature range" of phosphorescence discussed above. The high-temperature side of the TL peak has a larger half-width than the low-temperature side, which indicates that the TL peak is likely to consist of more than one TL peak overlapping with each other {\cite{mckeever1988thermoluminescence}}.

The solid black line in Fig {\ref{fig:GE TL with and without decay}} represents the TL glow curve recorded under the same conditions but with no delay time between the removal of excitation and the initiating of the temperature ramp. The integrated area of the whole TL glow curve depends on the total number of luminescence centres, regardless of whether they are spatially close to traps or not {\cite{dobrowolska2014electron}}. At low temperatures charge tunnelling occurs, and phosphorescence originates from donor-acceptor pairs. After this decay, there is still a significant concentration of neutral donors and acceptors that are physically isolated. It is not until the temperature is increased that these donors/acceptors are thermally ionized and release carriers that reset the close DAPs (luminescence centres), so they can emit multiple times and give rise to strong luminescence.

The thermoluminescence spectra (recorded after the phosphorescence has decayed) presents as one featureless broadband centred at 484 nm and 490 nm ($\sim$ 2.5 eV) for B\{001\} and C\{111\}, respectively. There is no shift in TL band position during heating for both samples.

\begin{figure}[!ht]
    \centering
    \includegraphics[scale=0.30]{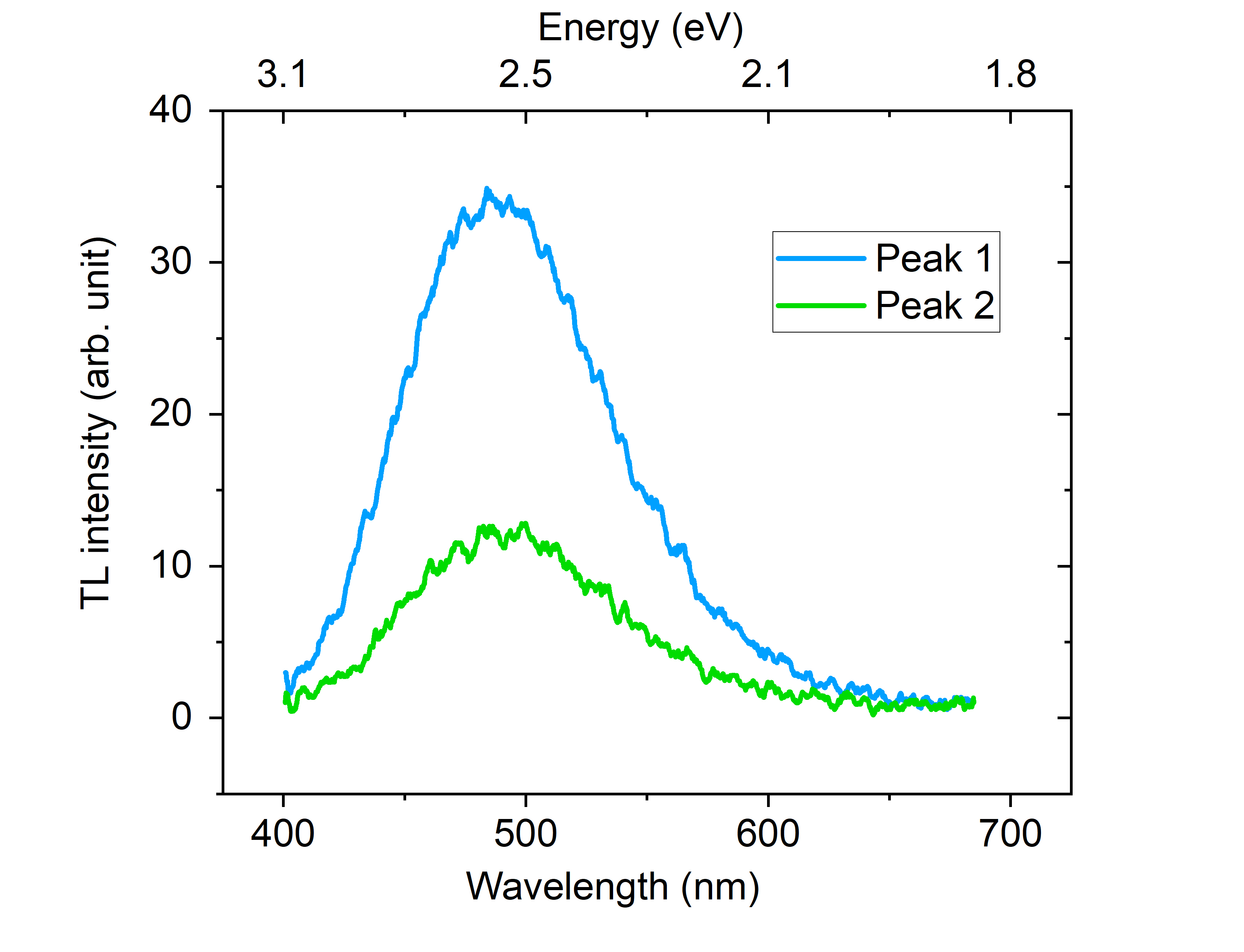}
    \caption{TL spectra at the peak temperature of ``Peak 1" (269~K) and ``Peak 2" (298 K) (after thermal cleaning) in sample C\{111\}.}
    \label{fig:GE81-107a-C TL clean spectra}
\end{figure}

By performing a ``TL cleaning" experiment [Fig~\ref{fig:TL experiment}], two independent TL glow peaks were obtained in both samples. When the temperature of peak intensity for ``Peak~1" was reached, the sample was cooled at \SI{50}{\kelvin\per\minute}, and the thermoluminescence intensity dropped rapidly to zero. The symmetry of ``Peak~2" observed during subsequent re-heating is typical of second-order kinetics (indicating that retrapping dominates) {\cite{mckeever1988thermoluminescence}}. Thus if the capture cross-section of trap and luminescence centre is equal, the number of luminescence centres is significantly less than that of traps. As shown in Fig {\ref{fig:GE81-107a-C TL clean spectra}}, the spectral position of maximum thermoluminescence emission in sample C\{111\} for ``Peak 1" and ``Peak 2" are both centred at 490 nm, which suggests free carriers thermally released from traps of different depth enable light emission from the same type of luminescence centres.

\section{Discussion}

\subsection{Determination of activation energies}

\begin{figure*}[!ht]
    \centering
    \includegraphics[scale=0.52]{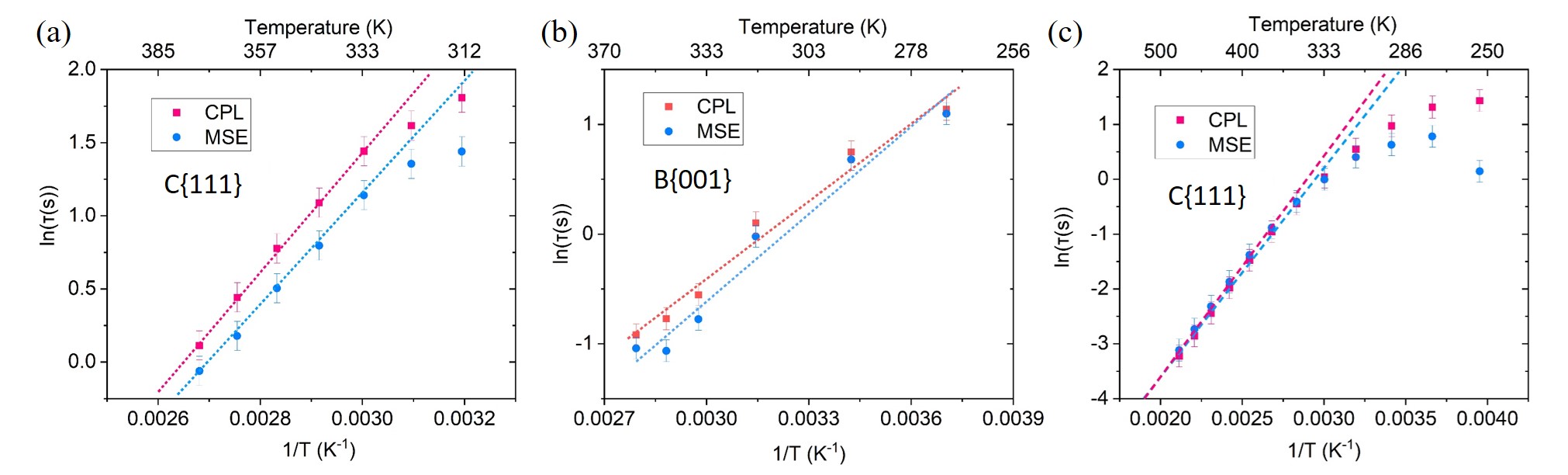}
    \caption{The logarithm of half-lifetime (fit by MSE function and CPL function, respectively) versus reciprocal temperature. The dash lines are the results of the fitting to an Arrhenius function. (a) integrated area of 2802 $\text{cm}^{-1}$ absorption peak (${\text{B}_\text{S}^0}$) decay in C\{111\} (b) integrated intensity of ${\text{N}_\text{S}^0}$ central EPR peak recovery in B\{001\} (c) phosphorescence decay in C\{111\} after 224 nm excitation.}
    \label{fig: MSE CPL EA fitting}
\end{figure*}

The post-illumination decay/recovery curves shown in Fig {\ref{fig:GE FTIR decay curves}}, {\ref{fig:GE EPR decay curves}} and {\ref{fig:GE phos decay different T}} can be fitted using different methods, including modified stretched exponential function (MSE) (see {\ref{apx:MSE}}), complex power-law function (CPL) (see {\ref{apx:CPL}}) {\cite{berberan2005mathematical,chen2003apparent,vedda2018tunneling,berberan2005mathematical,medlin1961decay,berberan2005mathematical2}}. In cases where multiple competing processes occur on similar timescales, a multi-component fit may be used to deduce the dominant decay routes (see {\ref{apx:multiple}}). In both samples, the multi-component fit indicates that in the high-temperature regime, thermal processes dominate and the decay is satisfactorily fit by two simple exponential components. By plotting the logarithm of half-lifetime versus reciprocal temperature (Fig {\ref{fig: MSE CPL EA fitting}}), the activation energy of $[\text{B}_\text{S}^0]$ decay (\SIrange{333}{373}{\kelvin}), $[\text{N}_\text{S}^0]$ decay (\SIrange{210}{250}{\kelvin})/recovery (\SIrange{270}{347}{\kelvin}), and phosphorescence decay (\SIrange{273}{473}{\kelvin}) can be determined (using the Arrhenius equation) and are listed in Table~{\ref{table: GE B activition energy summary}} and {\ref{table: GE C activition energy summary}}. 

The activation energies determined for $[\text{B}_\text{S}^0]$ decay in both samples are similar with an average value only slightly smaller than the ionization energy of the substitutional boron acceptor (0.37 eV {\cite{collins_71}}). 

The activation energies of the [$\text{N}_\text{S}^0$] recovery in B\{001\} determined by different methods are consistent and approximately 0.2 eV. In sample C\{111\}, measurements could not be made over a sufficiently wide temperature window, due to weak $\text{N}_\text{S}^0$ EPR signal strength and small optically induced changes in this signal, to accurately determine the activation energy for $[\text{N}_\text{S}^0]$ decay. 

At sufficiently high temperatures the activation energies for phosphorescence decays as determined by different fitting methods are in reasonable agreement. The average activation energy determined from sample C\{111\} is only slightly below that of the boron acceptor, while those from sample B\{001\} are significantly below suggesting that there is an important contribution from another trap with a significantly lower activation energy.

In the TL experiment, the depth of two traps were determined by the initial rise method and are listed in Table {\ref{table: GE B activition energy summary}} and {\ref{table: GE C activition energy summary}}. Accurate trap energies are difficult to determine for closely overlapping peaks, but the involvement of a trap with lower energy than the boron acceptor is suggested.

\begin{table*}[!ht]
\renewcommand\arraystretch{1.3}
\centering
\caption{Activation energies of phosphorescence decay, TL glow peaks, $\text{B}_\text{S}^0$ decay, and $\text{N}_\text{S}^0$ recovery after bandgap optical excitation in sample B\{001\} where $[\text{N}_\text{S}] > [\text{B}_\text{S}]$.}
\begin{ruledtabular}
\begin{tabular}{ccccc}
$E_A$ (eV) & \multicolumn{2}{c}{Multiple components} & MSE & CPL \\ 
 & Component 1 & Component 2 & & \\\hline
Phosphorescence & 0.21(5) & 0.23(5) & 0.30(5) & 0.25(5) \\
Thermoluminescence & 0.30(5)  & 0.33(5)  \\
$\text{B}_\text{S}^0$ decay & 0.32(5)  & 0.34(5) & 0.32(5) & 0.37(5) \\
$\text{N}_\text{S}^0$ recovery & 0.18(5)  & 0.21(5) & 0.22(5) & 0.21(5)  \\ 
\end{tabular}
\label{table: GE B activition energy summary}
\end{ruledtabular}
\end{table*}

\begin{table*}[!ht]
\renewcommand\arraystretch{1.3}
\centering
\caption{Activation energies of phosphorescence decay, TL glow peaks, $\text{B}_\text{S}^0$ decay, and $\text{N}_\text{S}^0$ decay after bandgap optical excitation in sample C\{111\} where $[\text{B}_\text{S}] > [\text{N}_\text{S}]$.}
\begin{ruledtabular}
\begin{tabular}{ccccc}
$E_A$ (eV) & \multicolumn{2}{c}{Multiple components} & MSE & CPL \\ 
 & Component 1 & Component 2 & & \\\hline
Phosphorescence & 0.31(5) & 0.33(5) & 0.34(5) & 0.35(5) \\
Thermoluminescence & 0.34(5)  & 0.37(5)  \\
$\text{B}_\text{S}^0$ decay & 0.34(5)  & 0.41(5) & 0.32(5) & 0.35(5)  \\
$\text{N}_\text{S}^0$ decay & $>$ 0.18  & $>$ 0.26 & $>$ 0.18 & $>$ 0.19 \\
\end{tabular}
\end{ruledtabular}
\label{table: GE C activition energy summary}
\end{table*}

\subsection{Charge transfer}

In sample B\{001\} where $[\text{N}_\text{S}] > [\text{B}_\text{S}]$, bandgap \SI{224}{\nano\meter} excitation decreases $[\text{N}_\text{S}]$, and when the illumination is removed the concentration recovers back to the value measured prior to bandgap illumination [Fig {\ref{fig:GE EPR decay curves}(b)}]. If the substitutional nitrogen defect can only exist in a neutral charge state and a positive charge state, the concentration of the neutral substitutional nitrogen is always expected to be increased (or not change) upon bandgap UV excitation in a sample containing only comparable quantities of substitutional nitrogen and boron defects (i.e., bandgap excitation creating free electron and holes will tend to neutralize donors and acceptors). However, the fact that this is not the case reveals that upon bandgap UV illumination another charge state of the substitutional nitrogen defect is produced. Therefore, it is proposed that the substitutional nitrogen defect can trap an electron to produce $\text{N}_\text{S}^-$ and this ``acceptor level" is only $\sim$ 0.2 eV below the conduction band minimum (Table {\ref{table: GE B activition energy summary}}). When $\text{N}_\text{S}^-$ is thermally ionized, the released electron can be trapped by $\text{N}_\text{S}^+$ and thus the concentration of $\text{N}_\text{S}^0$ can be increased. When boron acceptors are present after bandgap UV illumination is removed, DAP recombination will act to reduce the concentration of the neutral and negatively charged substitutional nitrogen defects. 

Several conclusions about the charge transfer processes can be drawn:
\begin{itemize}
 \item For the substitutional nitrogen defect three charge states have to be considered: negative, neutral, and positive. The total concentration of substitutional nitrogen is given by
 \begin{equation}
{[\text{N}_\text{T}]} = [\text{N}_\text{S}^-] + [\text{N}_\text{S}^0] + [\text{N}_\text{S}^+]  \label{eqn:total nitrogen conc}
\end{equation}
 \item For the substitutional boron defect, only the neutral and negatively charged states should be considered. The total concentration of substitutional boron is given by
 \begin{equation}
{[\text{B}_\text{T}]} = [\text{B}_\text{S}^-] + [\text{B}_\text{S}^0] \label{eqn:total boron conc}
\end{equation}

 \item Equilibrium charge balance is maintained such that after free carriers are trapped:
 \begin{equation}
[\text{N}_\text{S}^-] + \, [\text{B}_\text{S}^-] = [\text{N}_\text{S}^+]  \label{eqn:charge balance}
\end{equation}

  \item Near or above bandgap excitation generates pairs of free carriers or excitons. Free carrier recombination can happen through multiple relaxation channels including trapping. In diamond containing only substitutional nitrogen and boron impurities in low concentration, both tunneling between traps and thermal excitation of carriers from traps must be considered.
  
 \item Near or above bandgap excitation will increase the concentration of neutral substitutional boron acceptors:
 \begin{equation}
{\text{B}_\text{S}^-} + {\text{h}_\text{VB}^+} \to {\text{B}_\text{S}^0} 
\end{equation}
 \item Near or above bandgap excitation can either increase $[\text{N}_\text{S}^0]$
\begin{equation}
{\text{N}_\text{S}^+} + {\text{e}_\text{CB}^-} \to {\text{N}_\text{S}^0} 
\end{equation}
 or decrease $[\text{N}_\text{S}^0]$
 \begin{equation}
{\text{N}_\text{S}^0} + {\text{e}_\text{CB}^-} \to {\text{N}_\text{S}^-} 
\end{equation}
Optical excitation with sufficient (below-bandgap) energy to ionize $\text{N}_\text{S}^0$ can also decrease the concentration of $\text{N}_\text{S}^0$: 
 \begin{equation}
2{\text{N}_\text{S}^0} \overset{\hbar\omega}{\to} {\text{N}_\text{S}^+} + {\text{N}_\text{S}^-}  
\end{equation}

 \item $\text{N}_\text{S}^0$ is a deep donor and at the temperatures studied the probability of thermal ionization is negligible. However, $\text{N}_\text{S}^-$ is readily thermally ionized over a temperature range which overlaps with that for thermal ionization of $\text{B}_\text{S}^0$. 

\end{itemize}

\subsection{``Blue-green" phosphorescence mechanism}

The characteristic narrow lines associated with DAP emission from close pairs (separation of order or less than the sum of the donor and acceptor radii) that accompany the broad emission from distant pairs in other materials (e.g. GaP, 3C-SiC) with shallow, delocalised donors and acceptors are not observed in this work {\cite{hopfield1963pair,sun2011splitting}}. Given the compact donor and acceptor wavefunctions in diamond, there are very few neighbours that should even be considered as ``close pairs". Although Dischler \emph{et al.} tentatively interpreted CL emission features are arising from close DAP in CVD diamond, this assertion has not been confirmed and the intensities were not in accord with theory {\cite{dischler1994}}.

The identification of the ``blue-green" luminescence centre with $\text{N}_\text{S}^0$...$\text{B}_\text{S}^0$ DAP recombination is logical since $\text{N}_\text{S}$ and $\text{B}_\text{S}$ are the most abundant defects/impurities in near colourless HPHT diamond. This has previously been asserted {\cite{svcajev2013}} and this assignment had previously been erroneously ruled out or overlooked {\cite{eaton2011phosphorescence}}. The ``blue-green" emission observed in all the samples studied in this work would also be expected in CVD and natural diamonds co-doped with $\text{N}_\text{S}$ and $\text{B}_\text{S}$. 

\begin{figure}[!ht]
    \centering
    \includegraphics[scale=0.60]{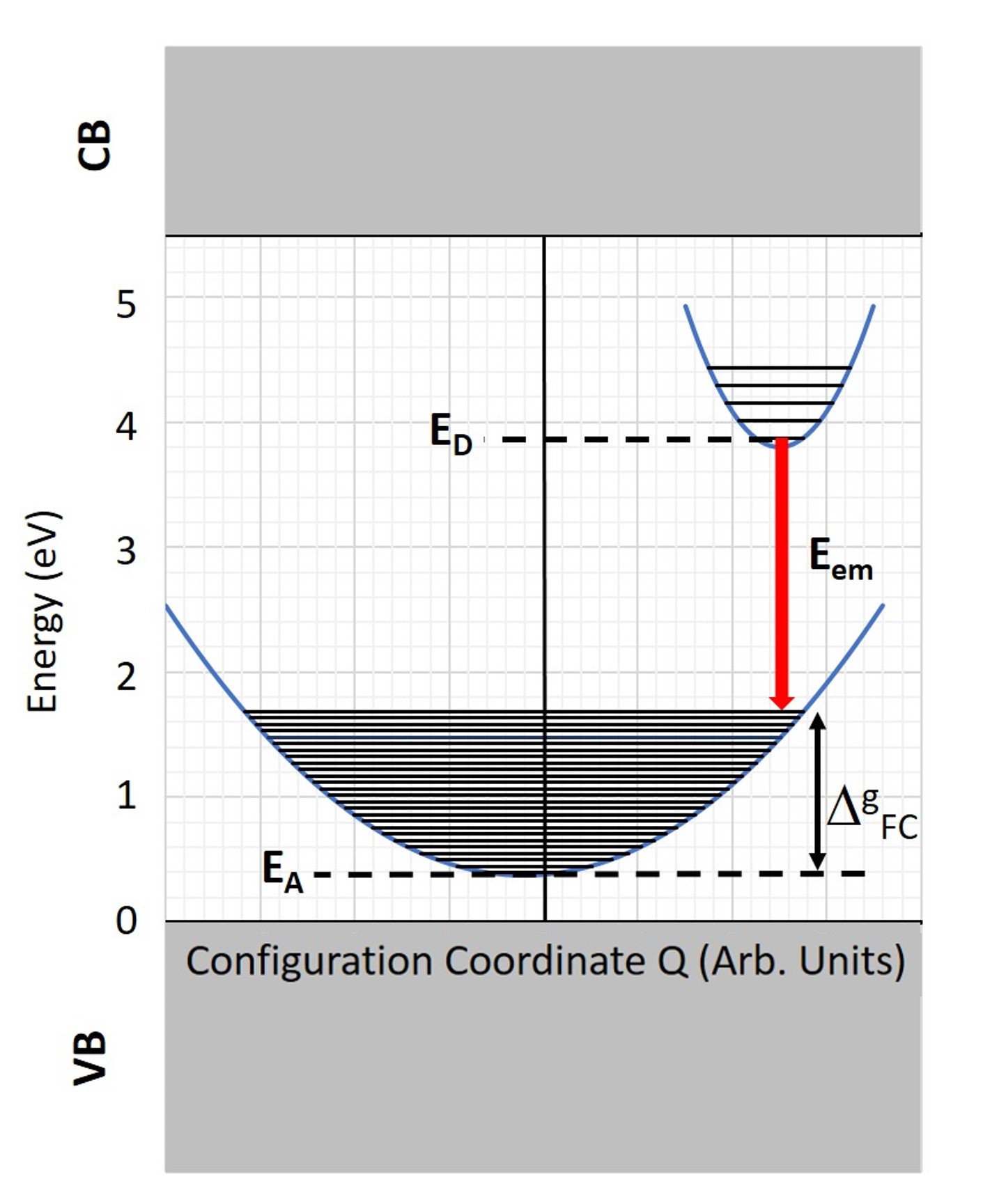}
    \caption{The configuration diagram for the energy levels of boron-nitrogen pair recombination at 0 K. The electronic state of substitutional nitrogen is 1.7 eV below the conduction band, and the ionisation energy of boron is 0.37 eV.}
    \label{fig:N and B recombination energy level}
\end{figure}

When modelling the donor-acceptor pair recombination process, the nitrogen donor changes charge state from $\text{N}_\text{S}^0$ to $\text{N}_\text{S}^+$ and there is a large configuration change which should not be ignored. ($\text{N}_\text{S}^0$ has $\text{C}_{\text{3}\text{v}}$ symmetry with one N-C bond 25 \% longer than the diamond C-C bond, and the other three smaller, whereas $\text{N}_\text{S}^+$ has $\text{T}_\text{d}$ symmetry with all N-C bonds approximately equal to the diamond C-C bond length {\cite{cox1994}}.) However, when the boron acceptor changes charge state from $\text{B}_\text{S}^0$ to $\text{B}_\text{S}^-$, there is little configurational change (in both charge states the defect has $\emph{T}_\text{d}$ symmetry; any symmetry lowering distortion is negligible {\cite{goss2006}}). It can be seen from a 1D configuration coordinate diagram [Fig {\ref{fig:N and B recombination energy level}}] that for a $\text{N}_\text{S}^0$...$\text{B}_\text{S}^0$ pair to emit a photon, a substantial amount of energy must be lost to the emission of a significant number of phonons. Fig {\ref{fig:N and B recombination energy level}} has been drawn assuming a typical $\text{N}_\text{S}^0$ phonon energy of 1130 cm$^{-1}$ ($\hbar\Omega_0$ = 141 meV){\cite{ashfold2020nitrogen}} and a $\text{B}_\text{S}^0$ phonon energy of 427 cm$^{-1}$ ($\hbar\Omega_0$ = 53 meV){\cite{mortet2018}} and that the energy lost to phonons in the configuration change associated with emission is approximately 1.2 eV. Taking this into account the energy of the DAP recombination transition energy should be given by 
\begin{equation}
\hbar\omega = E_g-(E_D + E_A) + \Delta E_C - \Delta^g_{FC} \label{eqn:DAP transition phonon}
\end{equation}
where $\Delta E_C$ is the Coulombic correction term and $\Delta^g_{FC}$ is the Frank-Condon shift. $\Delta^g_{FC}$ can be expressed as $SE_{ph}$, where the Huang-Rhys factor $S$ essentially qualifies the average number of phonons emitted per transition (in this case approximately 23) and $E_{ph}$ is the average phonon energy {\cite{alkauskas2016}}. Such strong electron-phonon coupling is unsurprising for such a deep localised donor. This is consistent with the description of N...B pair recombination emission hypothesised by $\text{\v{S}}{\text{\v{c}}}$ajev \emph{et al}. {\cite{svcajev2013}}

The Frank-Condon shift $\Delta^g_{FC}$ = 1.2 eV, would result in a distant $\text{N}_\text{S}^0$...$\text{B}_\text{S}^0$ pair emission energy of approximately 2.25 eV, consistent with the observations reported here and elsewhere. Given the large configuration change, such a large Frank-Condon shift appears reasonable, indeed in molecules the shifts can be as large as several eV {\cite{franceschetti2003}}. We note that the barrier to the reorientation of the $\text{N}_\text{S}^0$ $\emph{C}_{\text{3}\text{v}}$ axis has been estimated as approximately 0.8 eV {\cite{ammerlaan1981}}, and the energy of the $\text{N}_\text{S}^0$ donor level is deeper than the value predicted for a hydrogenic donor ($\sim$ 0.4 eV) by $\sim$ 1.3 eV. Both values are large compared to shifts observed in other materials and demonstrate the large configurational change associated with $\text{N}_\text{S}^0$. Further theoretical calculations of the Frank-Condon shift for emission from a $\text{N}_\text{S}^0$...$\text{B}_\text{S}^0$ pair would be very valuable.

\begin{figure}[!ht]
    \centering
    \includegraphics[scale=0.30]{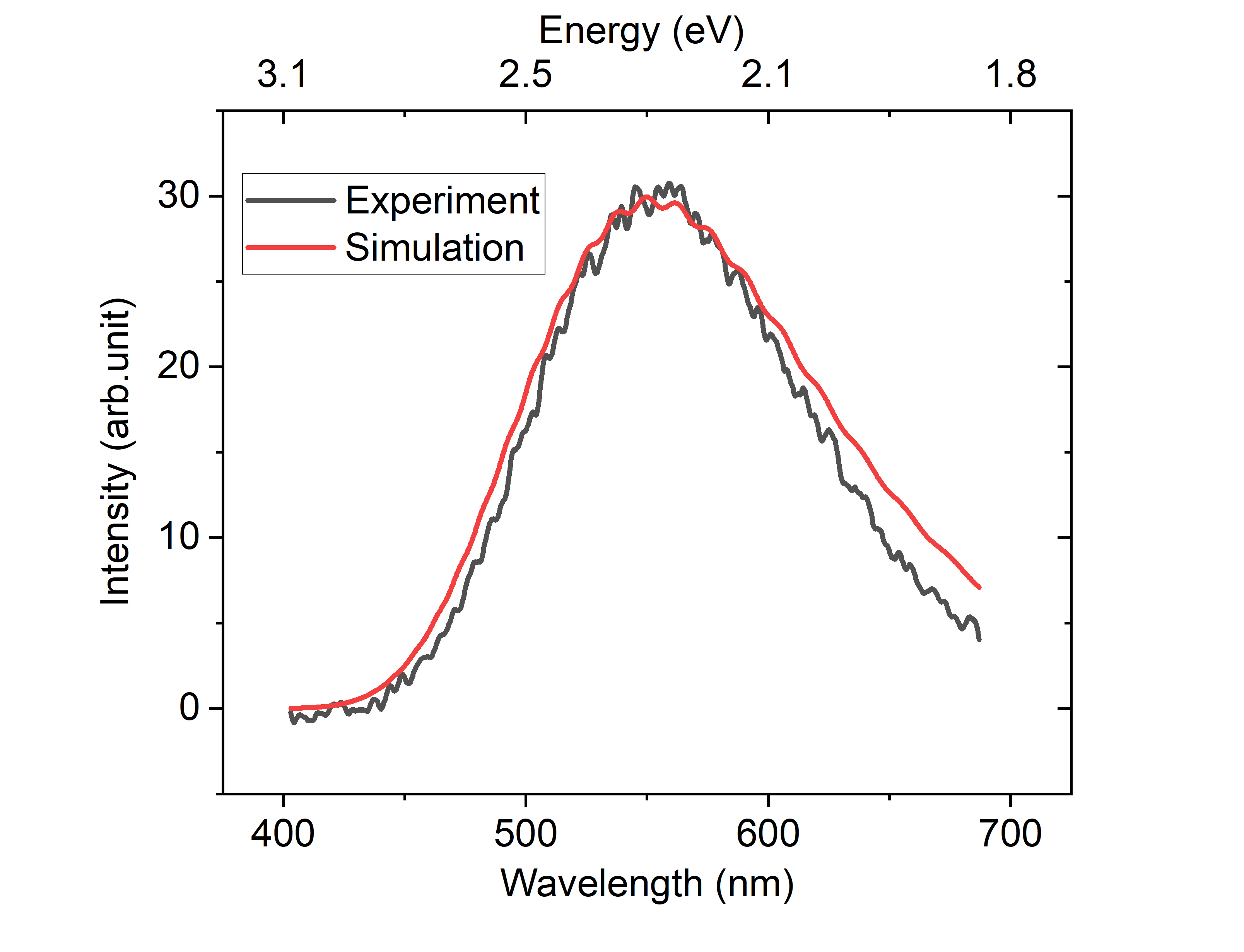}
    \caption[Comparison of experimental and simulated ``blue-green" phosphorescence spectrum]{Phosphorescence spectrum at 173 K in C\{111\} (black line), and simulated spectrum with parameters $E_g$ = 5.47 eV, $E_D$ = 1.7 eV ($\text{N}_\text{S}^0$), $E_A$ = 0.37 eV ($\text{B}_\text{S}^0$), $E_{ph} = \SI{53}{\milli\eV}$, $S$ = 22.7: $\Delta E_C$ is assumed zero (see text).}
    \label{fig:550 band simulation}
\end{figure}

To describe the vibrational broadening of the $\text{N}_\text{S}^0$...$\text{B}_\text{S}^0$ pair emission, it is necessary to sum all possible transitions between the accessible vibrational levels of the $\text{N}_\text{S}^0$ donor and all those of the associated $\text{B}_\text{S}^0$ acceptor. Assuming a typical $\text{N}_\text{S}^0$ phonon energy of $\hbar\Omega_0$ = 141 meV, then at room temperature ($k_B T$ $\sim$ 26 meV) only the ground vibrational state of the $\text{N}_\text{S}^0$ donor will be occupied, and thus using the 1D model it is only necessary to sum over all the vibrational levels of $\text{B}_\text{S}^0$ using equation
\begin{equation}
I(E) = \sum_n \, e^{-S} \, \frac{S^n}{n !} \, {g_\Gamma} \, (E_{ZPL} - nE_{ph} - E) \label{equ:phonon coupling lineshape}
\end{equation}
Where $n$ = 1,2,3... are the phonon replicas. $E_{ph}$ is the average phonon energy. $g_\Gamma$ is a Gaussian function with a parameter $\Gamma$ defining the width of the Gaussian peak. Strictly speaking this equation is derived assuming that the vibrational energy of the ground and excited state are equal but a similar result is obtained if they are not {\cite{alkauskas2012,alkauskas2016}}. For defects with large electron-phonon coupling ($S \gg 1$), resolved phonon replicas are not observed, and contributions from individual phonon modes cannot be identified. Furthermore, the luminescence intensity from the zero-phonon line of each different $\text{N}_\text{S}^0$...$\text{B}_\text{S}^0$ pair is practically zero. Fig {\ref{fig:550 band simulation}} shows a simulation of the distant (small Coulomb shift) $\text{N}_\text{S}^0$...$\text{B}_\text{S}^0$ pair emission in the limit of strong electron-phonon coupling ($S \sim 23$), using equation {\ref{equ:phonon coupling lineshape}}. The agreement between simple theory and experiment is reassuring. The broadening of the emission lines from the very few close pairs means that they are undetectable and the overlapping contributions from differently separated distant pairs (with different relaxation times) provides a natural explanation of the complex non-exponential decay.

\begin{figure}[!ht]
    \centering
    \includegraphics[scale=0.35]{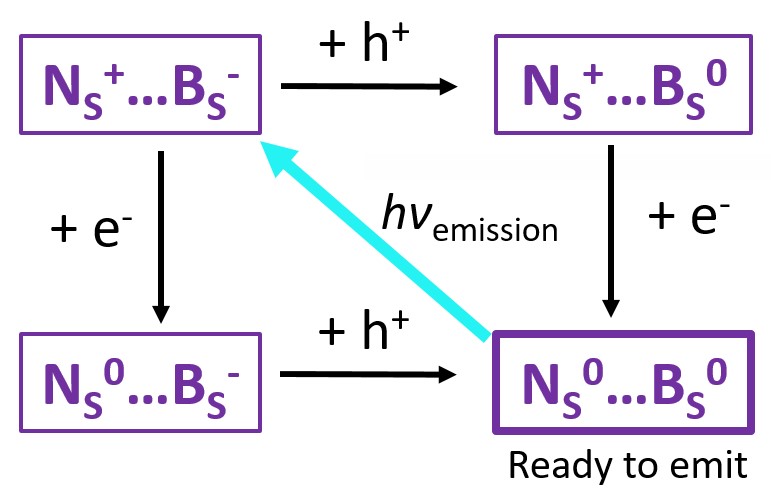}
    \caption{Schematic of the reset of ${\text{N}_\text{S}^0}...{\text{B}_\text{S}^0}$ luminescence centre.}
    \label{fig:NB paris luminescence centre}
\end{figure}

Before illumination, we expect a significant latent population of ${\text{N}_\text{S}^+}$ and ${\text{B}_\text{S}^-}$. In the same sample, significant populations of isolated ${\text{N}_\text{S}^0}$, ${\text{N}_\text{S}^-}$, and ${\text{B}_\text{S}^0}$ can be produced by near or above bandgap excitation. At low temperatures, there is essentially no thermal excitation of carriers to the conduction or valence band, and hence these defects are isolated, with only relatively close defects able to exchange charge via tunneling. Sufficiently close ${\text{N}_\text{S}^0}...{\text{B}_\text{S}^0}$ pairs can emit light:
\begin{equation}
{\text{N}_\text{S}^0}...{\text{B}_\text{S}^0} \to {\text{N}_\text{S}^+}...{\text{B}_\text{S}^-} + \hbar\omega 
\end{equation}
but only a very few have sufficiently close donors and acceptors to be reset into the ``ready to emit" ${\text{N}_\text{S}^0}...{\text{B}_\text{S}^0}$ state by tunneling. Thus, low temperature phosphorescence would be dominated by long-lived relatively distant ${\text{N}_\text{S}^0}...{\text{B}_\text{S}^0}$ pairs. Although their emission probability is typically low (long lifetime), there is a large population of them, and no other processes are operating to modify the charge states of these pairs' constituents. As the temperature is increased the probability of thermal ionization of ${\text{N}_\text{S}^-}$ and ${\text{B}_\text{S}^0}$ increases rapidly. Carriers are released into the conduction and valence bands, and defects are ``connected" by free carriers in these bands. Close ${\text{N}_\text{S}^0}...{\text{B}_\text{S}^0}$ pairs can emit, and may re-emit via the subsequent capture of both an electron and a hole. It is likely that distant ${\text{N}_\text{S}^0}...{\text{B}_\text{S}^0}$ pairs will have one or both constituents ionized before they emit. Thus at high temperature close pairs with higher energies will dominate the emission [Fig~ {\ref{fig:NB paris luminescence centre}}].

The energy at which a ${\text{N}_\text{S}^0}...{\text{B}_\text{S}^0}$ pair emits depends on the separation of the nitrogen donor and the boron acceptor, with the Coulomb correction term increasing as the separation between the donor and acceptor decreases. Statistically, there will be very few close pairs. The emission probability is higher for close pairs, thus the lifetime is shorter. 

Once a neutral donor-acceptor pair has emitted, both electron and hole capture (regardless of the order) are required to enable to pair the emit again as shown in figure {\ref{fig:NB paris luminescence centre}}. By resetting the charge states of the donor and the acceptor, it is possible for each luminescence centre to emit multiple times.

At low temperatures when there is no thermal activation of carriers, the emission will be dominated by distant pairs, and the majority of pairs will emit only once. Thus the peak emission will be at slightly lower energies (2.25 eV). As the temperature increases, the probability of thermal ionization increases and closer pairs can be reset by charge capture and emit multiple times. Furthermore, the long-lived distant pairs will be increasingly ionized before they can emit. Thus the emission peak will shift to higher energies (2.5 eV) as the temperature is increased. The magnitude of the shift observed is 0.25 eV, which is consistent with the Coulomb correction term. 

There is expected to only be a small configuration change between $\text{N}_\text{S}^0$ and $\text{N}_\text{S}^-$. The emission from a ${\text{N}_\text{S}^-}...{\text{B}_\text{S}^0}$ pair is expected to be in the UV range: if the Frank-Condon shift for this emission was $\sim$ \SIrange{1.2}{1.7}{\eV}, we estimate the DAP peak from ${\text{N}_\text{S}^-}...{\text{B}_\text{S}^0}$ pairs to be at $\sim$ \SIrange{3.2}{3.7}{\eV}.

\section{Conclusions}

Since the ``blue-green" phosphorescence spectrum, lifetime, and intensity vary between growth sectors in near colourless HPHT diamonds, it is important to study individual growth sectors in order to interpret the emission mechanism and defects involved. 

The data presented here confirms that the substitutional nitrogen defect in diamond has an acceptor state approximately 0.2 eV below the bottom of the conduction band. This is shallower than predicted by theory{\cite{jones2009acceptor}}, but interestingly such a low activation energy has been reported in previous studies,{\cite{walsh1971thermoluminescence, halperin1961some, halperin1966thermoluminescence, bourgoin1978thermally}} on natural and HPHT grown diamonds. 

It is shown that the broad ``blue-green" luminescence and phosphorescence band can be fully explained by emission from  $\text{N}_\text{S}^0$...$\text{B}_\text{S}^0$ donor-acceptor pairs, once the configurational change between charge states is considered. At sufficiently high temperatures $\text{N}_\text{S}^0$...$\text{B}_\text{S}^0$ luminescence pairs can be reset multiple times by capture of electrons and holes thermally ionized from $\text{N}_\text{S}^-$ and $\text{B}_\text{S}^0$, respectively. The ``blue-green" phosphorescence mechanism switches at low temperatures, where the probability of thermal ionization is negligible, to one where charge carrier tunneling dominates. The shift of the phosphorescence peak energy with increasing temperature and delay after the excitation is removed is consistent with the properties of the model proposed. 

The presence of both relatively shallow donors and acceptors is essential in order to reset the nitrogen-boron donor acceptor pairs into the ready-to-emit state ($\text{N}_\text{S}^0$...$\text{B}_\text{S}^0$) multiple times. This resetting of emitters facilitates bright and long-lived phosphorescence from a low concentration of relatively close donor-acceptor pairs. It must be remembered that in an insulating material like diamond the calculated position of the Fermi level does not necessarily predict the correct charge state of a defect, and defect charge state is strongly influenced by the proximity of a particular defect to a donor (or acceptor). After at or above bandgap excitation of near colourless diamond doped with relatively low concentrations of substitutional nitrogen and boron defects, at temperatures where the probability of thermal ionization is low, there can be significant populations of isolated $\text{N}_\text{S}^0$, $\text{B}_\text{S}^0$ and $\text{N}_\text{S}^-$ defects that persist almost indefinitely.   

This work predicts that UV luminescence should arise from $\text{N}_\text{S}^-$...$\text{B}_\text{S}^0$ donor acceptor pairs and that ``blue-green" luminescence and phosphorescence should be observed in diamond grown by chemical vapour deposition and co-doped with nitrogen and boron. In this work the concentration of $\text{N}_\text{S}^0$...$\text{B}_\text{S}^0$ pairs that are sufficiently close to give rise to optical emission has not been identified. It is usually assumed that the distribution of donors and acceptors will be random but it is not possible to rule out preferential incorporation of close pairs, or the production of pairs upon annealing and defect migration. Further work is required to address this question.


\begin{acknowledgments}
JZ thanks De Beers Ignite for providing funding. We thank the late Dr. Tom Anthony of the General Electric Research \& Development Centre for provision of samples used in this work. BLG gratefully acknowledges the Royal Academy of Engineering for a Research Fellowship. BLG and MEN acknowledge funding from EPSRC via grant EP/V056778/1.
\end{acknowledgments}

\appendix

\section{Decay/recovery curve fitting methods}

\subsection{Multiple components fitting}                 \label{apx:multiple}

The athermal tunnelling process does not involve the conduction band or the valence band but occurs between traps (donors or acceptors) and nearby luminescence centres with aligned energy levels in the bandgap \cite{vedda2018tunneling,avouris1981tunneling}. The probability of tunnelling is determined by the overlap of the trap/luminescence centre wave functions. Generally speaking, a high concentration of related defects enables the formation of close trap-centre pairs for tunnelling \cite{vedda2018tunneling}. The tunnelling process is also achievable between far distance trap and luminescence centre by a chain-like path which consists of several traps delivering carriers to the luminescence centre when the concentration of traps is high \cite{vedda2018tunneling}. Only considering pairs containing one variety of trap and luminescence centre separated by a distance $r$ (ignoring the chain-like path tunnelling), the phosphorescence intensity is given by
\begin{equation}
I(t) =  \frac{I_0}{(1 + t/\tau)}   \label{eqn:tunneling}
\end{equation}
where the lifetime $\tau$ depends on the intensity of excitation \cite{avouris1981trapping}.

For thermally activated process, electrons or holes are trapped at localized energy levels during optical excitation. After excitation, those trapped charge carriers can be thermally released into the conduction band or the valence band and then re-trapped at an isolated trap or radiative recombine at a centre, at which light may be subsequently emitted. The relative number of traps and luminescence centres determine whether retrapping or radiative recombination process dominates \cite{mckeever1988thermoluminescence,vedda2018tunneling}. When the concentration of luminescence centres is much higher than that of traps, recombination dominates, and the phosphorescence process follows first-order kinetics (only one variety of luminescence centres and one variety of traps are involved). The intensity of phosphorescence is described as 
\begin{equation}
I(t) =I_0\,\exp(-\frac{t}{\tau}), 
\end{equation}
where $I_0$ is the initial intensity, at very high temperature $I_0$ decreases. $t$ is a delay time, $\tau$ is the lifetime of traps:
\begin{equation}
\frac{1}{\tau} = \frac{1}{s} \, \exp(-\frac{E_{Trap}}{k_B T}), \label{equ:activation energy calculation}
\end{equation}
where $s$ is a constant, which may, however, vary slowly with temperature, $E_{Trap}$ is the depth of trap, $T$ is temperature. Therefore, for a constant $T$, the decay follows a simple exponential decay \cite{mckeever1988thermoluminescence,vedda2018tunneling}. When traps are more abundant, both retrapping and recombination are considered; if assume the initial number of trapped carriers and the initial number of luminescence centres available for recombination both equal to $n_0$, the phosphorescence process follows second-order kinetics. The intensity of phosphorescence is described as 
\begin{equation}
I(t) = \frac{I_{0}}{(n_{0}\,\alpha t + 1)^2}. \label{eqn:hyperbolic}
\end{equation}
The constant $\alpha$ describes the relative possibilities of the retrapping and the recombination processes. The phosphorescence decay follows a power-law decay with the power of two, namely a hyperbolic decay \cite{mckeever1988thermoluminescence}.

Multiple varieties of traps and luminescence centres may participate in the thermal and tunnelling processes. The phosphorescence decay, therefore, could be considered as the combination of multiple decay components. Accordingly, the intensity of phosphorescence is described by
\begin{equation}
I(t) = \sum_m\,\frac{I_0}{(1 + t/\tau)} + \sum_n\,I_{0}\,\exp(-\frac{t}{\tau}) + \sum_p\,\frac{I_{0}}{(n_{0}\,\alpha t + 1)^2} \label{equ:multiple components phos function}
\end{equation}
where $m$, $n$, $p$ are integers $\ge$ 0, representing the number of components belongs to 1/t power law decay, simple exponential decay, and hyperbolic decay in phosphorescence, respectively. 

\subsection{Stretched exponential function}              \label{apx:MSE}

The stretched exponential function, as a sum of exponential decay, is robust to describe complex thermally activated luminescence decay considering the diversity of impurities and charge transfer paths, as well as the random distribution traps \cite{berberan2005mathematical, phillips1996}. During phosphorescence decay, the number of available luminescence centres reduce, thereby free carriers in the conduction band or the valence band must travel for a longer distance before activating recombination at luminescence centres, leading to a decrease in the phosphorescence decay rate. The stretched exponential function described such a luminescence decay without a constant decay rate with the assumptions of one at least one trapping state and one variety of luminescence centre involved, direct band-to-band recombination is negligible, and retrapping is considered \cite{chen2003apparent}. The intensity of phosphorescence decay described by a stretched exponential function, which is also called Kohlrausch function, is written as
\begin{equation}
I(t) = I_0\,\exp[-(\frac{t}{\tau_0})^\beta], 
\end{equation}
where $I_0$ is the initial intensity, $t$ is a delay time, $\tau_0$ is a parameter with the dimensions of time, $\beta$ is a dispersion exponent with a range of $0 < \beta < 1$ \cite{berberan2005mathematical,chen2003apparent}. The time-dependent rate coefficient $k(t)$ is 
\begin{equation}
k(t) = \frac{\beta}{\tau_0}\,(\frac{t}{\tau_0})^{\beta-1}. 
\end{equation}
When $k(t)$ is constant, the decay is an exponential decay. When $k(t)$ increases or decreases with time, the decay described is a super-exponential decay or a sub-exponential decay, respectively \cite{berberan2005mathematical}.

The infinite initial decay rate of the stretched exponential function is obviously not in accordance with the understanding of phosphorescence decay. To solve the problem of fitting the first several experimental data points of the decay curve, a modified stretched exponential (MSE) function is created to enable the selection of the origin of times $t_0$ comparing with $\tau_0$. A parameter $\alpha$ is defined as $t_0/\tau_0$. Hence, the modified exponential function for phosphorescence decay is 
\begin{equation}
I(t) = \exp\,[\alpha^\beta - (\alpha + \frac{t}{\tau_0})^\beta]. \label{equ:modified stretched exponential function}
\end{equation}
The time-dependent rate coefficient is 
\begin{equation}
k(t) = \frac{\beta}{\tau_0}\,(\alpha + \frac{t}{\tau_0})^{\beta - 1}. 
\end{equation}
The lifetime of phosphorescence is given by 
\begin{equation}
\tau = \tau_0\,\frac{\tau^{1-\beta}}{\beta}. 
\end{equation}
In summary, a better approach of experimental fits for phosphorescence decay of complex thermal processes is provided by a modified stretched exponential function \cite{vedda2018tunneling}. Whilst the modified stretched exponential function is not a complete model of phosphorescence decay, it does achieve better accuracy in calculating the activation energy.

\subsection{Complex power-law (CPL) function}           \label{apx:CPL}

A complex power-law function (compressed hyperbola) was devised by Becquerel to provide a more general solution for phosphorescence decay fitting as a supplement to the exponential function and to accommodate more possible factors, such as the distribution of traps with different depth and the interaction of different luminescence centres. The equation is written as
\begin{equation}
I(t) = \frac{I_0}{(1 + at)^p}, \label{equ:complex power law function}
\end{equation}
where $I_0$ is initial intensity, $a$ is a constant, and $p$ is the power between 1 - 2 \cite{berberan2005mathematical2,medlin1961decay}.

\nocite{*}

\bibliographystyle{apsrev4-2}
\bibliography{apssamp}

\begin{thebibliography}{65}%
\makeatletter
\providecommand \@ifxundefined [1]{%
 \@ifx{#1\undefined}
}%
\providecommand \@ifnum [1]{%
 \ifnum #1\expandafter \@firstoftwo
 \else \expandafter \@secondoftwo
 \fi
}%
\providecommand \@ifx [1]{%
 \ifx #1\expandafter \@firstoftwo
 \else \expandafter \@secondoftwo
 \fi
}%
\providecommand \natexlab [1]{#1}%
\providecommand \enquote  [1]{``#1''}%
\providecommand \bibnamefont  [1]{#1}%
\providecommand \bibfnamefont [1]{#1}%
\providecommand \citenamefont [1]{#1}%
\providecommand \href@noop [0]{\@secondoftwo}%
\providecommand \href [0]{\begingroup \@sanitize@url \@href}%
\providecommand \@href[1]{\@@startlink{#1}\@@href}%
\providecommand \@@href[1]{\endgroup#1\@@endlink}%
\providecommand \@sanitize@url [0]{\catcode `\\12\catcode `\$12\catcode
  `\&12\catcode `\#12\catcode `\^12\catcode `\_12\catcode `\%12\relax}%
\providecommand \@@startlink[1]{}%
\providecommand \@@endlink[0]{}%
\providecommand \url  [0]{\begingroup\@sanitize@url \@url }%
\providecommand \@url [1]{\endgroup\@href {#1}{\urlprefix }}%
\providecommand \urlprefix  [0]{URL }%
\providecommand \Eprint [0]{\href }%
\providecommand \doibase [0]{https://doi.org/}%
\providecommand \selectlanguage [0]{\@gobble}%
\providecommand \bibinfo  [0]{\@secondoftwo}%
\providecommand \bibfield  [0]{\@secondoftwo}%
\providecommand \translation [1]{[#1]}%
\providecommand \BibitemOpen [0]{}%
\providecommand \bibitemStop [0]{}%
\providecommand \bibitemNoStop [0]{.\EOS\space}%
\providecommand \EOS [0]{\spacefactor3000\relax}%
\providecommand \BibitemShut  [1]{\csname bibitem#1\endcsname}%
\let\auto@bib@innerbib\@empty
\bibitem [{\citenamefont {Farrer}(1969)}]{farrer1969substitutional}%
  \BibitemOpen
  \bibfield  {author} {\bibinfo {author} {\bibfnamefont {R.~G.}\ \bibnamefont
  {Farrer}},\ }\href
  {https://www.sciencedirect.com/science/article/abs/pii/0038109869905936}
  {\bibfield  {journal} {\bibinfo  {journal} {Solid State Commun.}\ }\textbf
  {\bibinfo {volume} {7}},\ \bibinfo {pages} {685} (\bibinfo {year}
  {1969})}\BibitemShut {NoStop}%
\bibitem [{\citenamefont {Stenger}\ \emph {et~al.}(2013)\citenamefont
  {Stenger}, \citenamefont {Pinault-Thaury}, \citenamefont {Kociniewski},
  \citenamefont {Lusson}, \citenamefont {Chikoidze}, \citenamefont {Jomard},
  \citenamefont {Dumont}, \citenamefont {Chevallier},\ and\ \citenamefont
  {Barjon}}]{Stenger2013}%
  \BibitemOpen
  \bibfield  {author} {\bibinfo {author} {\bibfnamefont {I.}~\bibnamefont
  {Stenger}}, \bibinfo {author} {\bibfnamefont {M.-A.}\ \bibnamefont
  {Pinault-Thaury}}, \bibinfo {author} {\bibfnamefont {T.}~\bibnamefont
  {Kociniewski}}, \bibinfo {author} {\bibfnamefont {A.}~\bibnamefont {Lusson}},
  \bibinfo {author} {\bibfnamefont {E.}~\bibnamefont {Chikoidze}}, \bibinfo
  {author} {\bibfnamefont {F.}~\bibnamefont {Jomard}}, \bibinfo {author}
  {\bibfnamefont {Y.}~\bibnamefont {Dumont}}, \bibinfo {author} {\bibfnamefont
  {J.}~\bibnamefont {Chevallier}},\ and\ \bibinfo {author} {\bibfnamefont
  {J.}~\bibnamefont {Barjon}},\ }\href {https://doi.org/10.1063/1.4818946}
  {\bibfield  {journal} {\bibinfo  {journal} {Journal of Applied Physics}\
  }\textbf {\bibinfo {volume} {114}},\ \bibinfo {pages} {073711} (\bibinfo
  {year} {2013})}\BibitemShut {NoStop}%
\bibitem [{\citenamefont {COLLINS}(1993)}]{collins_71}%
  \BibitemOpen
  \bibfield  {author} {\bibinfo {author} {\bibfnamefont {A.}~\bibnamefont
  {COLLINS}},\ }\href {https://doi.org/10.1098/rsta.1993.0017} {\bibfield
  {journal} {\bibinfo  {journal} {Philosophical Transactions of the Royal
  Society of London Series A-Mathematical Physical and Engineering Sciences}\
  }\textbf {\bibinfo {volume} {342}},\ \bibinfo {pages} {233} (\bibinfo {year}
  {1993})}\BibitemShut {NoStop}%
\bibitem [{\citenamefont {Sonin}\ \emph {et~al.}(2022)\citenamefont {Sonin},
  \citenamefont {Tomilenko}, \citenamefont {Zhimulev}, \citenamefont {Bul'ak},
  \citenamefont {Chepurov}, \citenamefont {Babich}, \citenamefont {Logvinova},
  \citenamefont {Timina},\ and\ \citenamefont {Chepurov}}]{Sonin2022}%
  \BibitemOpen
  \bibfield  {author} {\bibinfo {author} {\bibfnamefont {V.}~\bibnamefont
  {Sonin}}, \bibinfo {author} {\bibfnamefont {A.}~\bibnamefont {Tomilenko}},
  \bibinfo {author} {\bibfnamefont {E.}~\bibnamefont {Zhimulev}}, \bibinfo
  {author} {\bibfnamefont {T.}~\bibnamefont {Bul'ak}}, \bibinfo {author}
  {\bibfnamefont {A.}~\bibnamefont {Chepurov}}, \bibinfo {author}
  {\bibfnamefont {Y.}~\bibnamefont {Babich}}, \bibinfo {author} {\bibfnamefont
  {A.}~\bibnamefont {Logvinova}}, \bibinfo {author} {\bibfnamefont
  {T.}~\bibnamefont {Timina}},\ and\ \bibinfo {author} {\bibfnamefont
  {A.}~\bibnamefont {Chepurov}},\ }\href
  {https://doi.org/10.1038/s41598-022-05153-7} {\bibfield  {journal} {\bibinfo
  {journal} {Scientific Reports}\ }\textbf {\bibinfo {volume} {12}},\ \bibinfo
  {pages} {1246} (\bibinfo {year} {2022})}\BibitemShut {NoStop}%
\bibitem [{Note1()}]{Note1}%
  \BibitemOpen
  \bibinfo {note} {Type II: diamond contains very low level of nitrogen
  impurities ($<$ 1 ppm)}\BibitemShut {NoStop}%
\bibitem [{\citenamefont {Watanabe}\ \emph {et~al.}(1997)\citenamefont
  {Watanabe}, \citenamefont {Lawson}, \citenamefont {Isoya}, \citenamefont
  {Kanda},\ and\ \citenamefont {Sato}}]{watanabe1997phosphorescence}%
  \BibitemOpen
  \bibfield  {author} {\bibinfo {author} {\bibfnamefont {K.}~\bibnamefont
  {Watanabe}}, \bibinfo {author} {\bibfnamefont {S.~C.}\ \bibnamefont
  {Lawson}}, \bibinfo {author} {\bibfnamefont {J.}~\bibnamefont {Isoya}},
  \bibinfo {author} {\bibfnamefont {H.}~\bibnamefont {Kanda}},\ and\ \bibinfo
  {author} {\bibfnamefont {Y.}~\bibnamefont {Sato}},\ }\href
  {https://www.sciencedirect.com/science/article/abs/pii/S0925963596007649}
  {\bibfield  {journal} {\bibinfo  {journal} {Diamond and Related Materials}\
  }\textbf {\bibinfo {volume} {6}},\ \bibinfo {pages} {99} (\bibinfo {year}
  {1997})}\BibitemShut {NoStop}%
\bibitem [{\citenamefont {Eaton-Maga{\~n}a}\ and\ \citenamefont
  {Breeding}(2016)}]{eaton2016}%
  \BibitemOpen
  \bibfield  {author} {\bibinfo {author} {\bibfnamefont {S.}~\bibnamefont
  {Eaton-Maga{\~n}a}}\ and\ \bibinfo {author} {\bibfnamefont {C.~M.}\
  \bibnamefont {Breeding}},\ }\href
  {https://static1.squarespace.com/static/56c5ef57859fd08f9874b6cc/t/60957f2ba6a0a346afd74013/1620410157967/2016+Eaton-Magana+and+Breeding%2C+An+Intro+to+PL+spec+for+diamond+and+apps+in+gemology%2C+Gems+%26+Gemology.pdf}
  {\bibfield  {journal} {\bibinfo  {journal} {Gems \& Gemology}\ }\textbf
  {\bibinfo {volume} {52}},\ \bibinfo {pages} {2} (\bibinfo {year}
  {2016})}\BibitemShut {NoStop}%
\bibitem [{\citenamefont {D'Haenens-Johansson}\ \emph
  {et~al.}(2015)\citenamefont {D'Haenens-Johansson}, \citenamefont {Katrusha},
  \citenamefont {Moe}, \citenamefont {Johnson},\ and\ \citenamefont
  {Wang}}]{Ulrika2015large}%
  \BibitemOpen
  \bibfield  {author} {\bibinfo {author} {\bibfnamefont {U.~F.~S.}\
  \bibnamefont {D'Haenens-Johansson}}, \bibinfo {author} {\bibfnamefont
  {A.}~\bibnamefont {Katrusha}}, \bibinfo {author} {\bibfnamefont {K.~S.}\
  \bibnamefont {Moe}}, \bibinfo {author} {\bibfnamefont {P.}~\bibnamefont
  {Johnson}},\ and\ \bibinfo {author} {\bibfnamefont {W.}~\bibnamefont
  {Wang}},\ }\href
  {https://www.gia.edu/gems-gemology/fall-2015-large-colorless-hpht-grown-synthetic-gem-diamond-technology-russia}
  {\bibfield  {journal} {\bibinfo  {journal} {Gems \& Gemology}\ }\textbf
  {\bibinfo {volume} {51}},\ \bibinfo {pages} {260} (\bibinfo {year}
  {2015})}\BibitemShut {NoStop}%
\bibitem [{\citenamefont {Eaton-Maga{\~n}a}\ and\ \citenamefont
  {Lu}(2011)}]{eaton2011phosphorescence}%
  \BibitemOpen
  \bibfield  {author} {\bibinfo {author} {\bibfnamefont {S.}~\bibnamefont
  {Eaton-Maga{\~n}a}}\ and\ \bibinfo {author} {\bibfnamefont {R.}~\bibnamefont
  {Lu}},\ }\href
  {https://www.sciencedirect.com/science/article/abs/pii/S0925963511001713}
  {\bibfield  {journal} {\bibinfo  {journal} {Diamond and Related Materials}\
  }\textbf {\bibinfo {volume} {20}},\ \bibinfo {pages} {983} (\bibinfo {year}
  {2011})}\BibitemShut {NoStop}%
\bibitem [{\citenamefont {Eaton-Maga{\~n}a}\ \emph {et~al.}(2008)\citenamefont
  {Eaton-Maga{\~n}a}, \citenamefont {Post}, \citenamefont {Heaney},
  \citenamefont {Freitas}, \citenamefont {Klein}, \citenamefont {Walters},\
  and\ \citenamefont {Butler}}]{eaton2008using}%
  \BibitemOpen
  \bibfield  {author} {\bibinfo {author} {\bibfnamefont {S.}~\bibnamefont
  {Eaton-Maga{\~n}a}}, \bibinfo {author} {\bibfnamefont {J.~E.}\ \bibnamefont
  {Post}}, \bibinfo {author} {\bibfnamefont {P.~J.}\ \bibnamefont {Heaney}},
  \bibinfo {author} {\bibfnamefont {J.}~\bibnamefont {Freitas}}, \bibinfo
  {author} {\bibfnamefont {P.}~\bibnamefont {Klein}}, \bibinfo {author}
  {\bibfnamefont {R.}~\bibnamefont {Walters}},\ and\ \bibinfo {author}
  {\bibfnamefont {J.~E.}\ \bibnamefont {Butler}},\ }\href
  {https://pubs.geoscienceworld.org/gsa/geology/article-abstract/36/1/83/130015/Using-phosphorescence-as-a-fingerprint-for-the?redirectedFrom=fulltext}
  {\bibfield  {journal} {\bibinfo  {journal} {Geology}\ }\textbf {\bibinfo
  {volume} {36}},\ \bibinfo {pages} {83} (\bibinfo {year} {2008})}\BibitemShut
  {NoStop}%
\bibitem [{\citenamefont {Walsh}\ \emph {et~al.}(1971)\citenamefont {Walsh},
  \citenamefont {Lightowlers},\ and\ \citenamefont
  {Collins}}]{walsh1971thermoluminescence}%
  \BibitemOpen
  \bibfield  {author} {\bibinfo {author} {\bibfnamefont {P.~S.}\ \bibnamefont
  {Walsh}}, \bibinfo {author} {\bibfnamefont {E.~C.}\ \bibnamefont
  {Lightowlers}},\ and\ \bibinfo {author} {\bibfnamefont {A.~T.}\ \bibnamefont
  {Collins}},\ }\href
  {https://www.sciencedirect.com/science/article/abs/pii/0022231371900391}
  {\bibfield  {journal} {\bibinfo  {journal} {J. Lumin}\ }\textbf {\bibinfo
  {volume} {4}},\ \bibinfo {pages} {369} (\bibinfo {year} {1971})}\BibitemShut
  {NoStop}%
\bibitem [{\citenamefont {Song}\ \emph {et~al.}(2016)\citenamefont {Song},
  \citenamefont {Lu}, \citenamefont {Tang}, \citenamefont {Ke}, \citenamefont
  {Su}, \citenamefont {Gao}, \citenamefont {Hu}, \citenamefont {Zhang},
  \citenamefont {Zhou}, \citenamefont {Bi} \emph
  {et~al.}}]{song2016identification}%
  \BibitemOpen
  \bibfield  {author} {\bibinfo {author} {\bibfnamefont {Z.}~\bibnamefont
  {Song}}, \bibinfo {author} {\bibfnamefont {T.}~\bibnamefont {Lu}}, \bibinfo
  {author} {\bibfnamefont {S.}~\bibnamefont {Tang}}, \bibinfo {author}
  {\bibfnamefont {J.}~\bibnamefont {Ke}}, \bibinfo {author} {\bibfnamefont
  {J.}~\bibnamefont {Su}}, \bibinfo {author} {\bibfnamefont {B.}~\bibnamefont
  {Gao}}, \bibinfo {author} {\bibfnamefont {N.}~\bibnamefont {Hu}}, \bibinfo
  {author} {\bibfnamefont {J.}~\bibnamefont {Zhang}}, \bibinfo {author}
  {\bibfnamefont {J.}~\bibnamefont {Zhou}}, \bibinfo {author} {\bibfnamefont
  {L.}~\bibnamefont {Bi}}, \emph {et~al.},\ }\href
  {https://www.researchgate.net/profile/Taijin-Lu/publication/305223678_Identification_of_Colourless_HPHT-grown_Synthetic_Diamonds_from_Shandong_China/links/57859ec308aec5c2e4e12084/Identification-of-Colourless-HPHT-grown-Synthetic-Diamonds-from-Shandong-China.pdf}
  {\bibfield  {journal} {\bibinfo  {journal} {J. Gemmol}\ }\textbf {\bibinfo
  {volume} {35}},\ \bibinfo {pages} {140} (\bibinfo {year} {2016})}\BibitemShut
  {NoStop}%
\bibitem [{\citenamefont
  {Chandrasekharan}(1946{\natexlab{a}})}]{chandrasekharan1946patterns}%
  \BibitemOpen
  \bibfield  {author} {\bibinfo {author} {\bibfnamefont {V.}~\bibnamefont
  {Chandrasekharan}},\ }in\ \href@noop {} {\emph {\bibinfo {booktitle}
  {Proceedings of the Indian Academy of Sciences-Section A}}},\ Vol.~\bibinfo
  {volume} {24}\ (\bibinfo {organization} {Springer India},\ \bibinfo {year}
  {1946})\ pp.\ \bibinfo {pages} {182--186}\BibitemShut {NoStop}%
\bibitem [{\citenamefont
  {Chandrasekharan}(1946{\natexlab{b}})}]{chandrasekharan1946TL}%
  \BibitemOpen
  \bibfield  {author} {\bibinfo {author} {\bibfnamefont {V.}~\bibnamefont
  {Chandrasekharan}},\ }in\ \href@noop {} {\emph {\bibinfo {booktitle}
  {Proceedings of the Indian Academy of Sciences-Section A}}},\ Vol.~\bibinfo
  {volume} {24}\ (\bibinfo {organization} {Springer India},\ \bibinfo {year}
  {1946})\ p.\ \bibinfo {pages} {187}\BibitemShut {NoStop}%
\bibitem [{\citenamefont {Krumme}\ and\ \citenamefont
  {Leivo}(1964)}]{krumme1964}%
  \BibitemOpen
  \bibfield  {author} {\bibinfo {author} {\bibfnamefont {J.~B.}\ \bibnamefont
  {Krumme}}\ and\ \bibinfo {author} {\bibfnamefont {W.~J.}\ \bibnamefont
  {Leivo}},\ }in\ \href@noop {} {\emph {\bibinfo {booktitle} {Proceedings of
  the Oklahoma Academy of Science}}}\ (\bibinfo {year} {1964})\ pp.\ \bibinfo
  {pages} {105--114}\BibitemShut {NoStop}%
\bibitem [{\citenamefont {Su}\ \emph {et~al.}(2018)\citenamefont {Su},
  \citenamefont {Zhao}, \citenamefont {Lou}, \citenamefont {Niu}, \citenamefont
  {Fang}, \citenamefont {Li}, \citenamefont {Shen}, \citenamefont {Zang},
  \citenamefont {Jia},\ and\ \citenamefont {Shan}}]{su2018}%
  \BibitemOpen
  \bibfield  {author} {\bibinfo {author} {\bibfnamefont {L.}~\bibnamefont
  {Su}}, \bibinfo {author} {\bibfnamefont {C.}~\bibnamefont {Zhao}}, \bibinfo
  {author} {\bibfnamefont {Q.}~\bibnamefont {Lou}}, \bibinfo {author}
  {\bibfnamefont {C.}~\bibnamefont {Niu}}, \bibinfo {author} {\bibfnamefont
  {C.}~\bibnamefont {Fang}}, \bibinfo {author} {\bibfnamefont {Z.}~\bibnamefont
  {Li}}, \bibinfo {author} {\bibfnamefont {C.}~\bibnamefont {Shen}}, \bibinfo
  {author} {\bibfnamefont {J.}~\bibnamefont {Zang}}, \bibinfo {author}
  {\bibfnamefont {X.}~\bibnamefont {Jia}},\ and\ \bibinfo {author}
  {\bibfnamefont {C.}~\bibnamefont {Shan}},\ }\href
  {https://www.sciencedirect.com/science/article/abs/pii/S0008622318300459}
  {\bibfield  {journal} {\bibinfo  {journal} {Carbon}\ }\textbf {\bibinfo
  {volume} {130}},\ \bibinfo {pages} {384} (\bibinfo {year}
  {2018})}\BibitemShut {NoStop}%
\bibitem [{\citenamefont {McKeever}(1988)}]{mckeever1988thermoluminescence}%
  \BibitemOpen
  \bibfield  {author} {\bibinfo {author} {\bibfnamefont {S.~W.~S.}\
  \bibnamefont {McKeever}},\ }\href@noop {} {\emph {\bibinfo {title}
  {Thermoluminescence of Solids}}},\ Vol.~\bibinfo {volume} {3}\ (\bibinfo
  {publisher} {Cambridge University Press},\ \bibinfo {year}
  {1988})\BibitemShut {NoStop}%
\bibitem [{\citenamefont {Paslovsky}\ \emph {et~al.}(1993)\citenamefont
  {Paslovsky}, \citenamefont {Lowther}, \citenamefont {Nam},\ and\
  \citenamefont {Keddy}}]{paslovsky1993interpretation}%
  \BibitemOpen
  \bibfield  {author} {\bibinfo {author} {\bibfnamefont {L.}~\bibnamefont
  {Paslovsky}}, \bibinfo {author} {\bibfnamefont {J.~E.}\ \bibnamefont
  {Lowther}}, \bibinfo {author} {\bibfnamefont {T.~L.}\ \bibnamefont {Nam}},\
  and\ \bibinfo {author} {\bibfnamefont {R.~J.}\ \bibnamefont {Keddy}},\ }\href
  {https://www.sciencedirect.com/science/article/abs/pii/002223139390038O}
  {\bibfield  {journal} {\bibinfo  {journal} {J. Lumin}\ }\textbf {\bibinfo
  {volume} {55}},\ \bibinfo {pages} {167} (\bibinfo {year} {1993})}\BibitemShut
  {NoStop}%
\bibitem [{\citenamefont {Petitfils}\ \emph {et~al.}(2007)\citenamefont
  {Petitfils}, \citenamefont {Wrobel}, \citenamefont {Benabdesselam},
  \citenamefont {Iacconi},\ and\ \citenamefont {Butler}}]{petitfils2007role}%
  \BibitemOpen
  \bibfield  {author} {\bibinfo {author} {\bibfnamefont {A.}~\bibnamefont
  {Petitfils}}, \bibinfo {author} {\bibfnamefont {F.}~\bibnamefont {Wrobel}},
  \bibinfo {author} {\bibfnamefont {M.}~\bibnamefont {Benabdesselam}}, \bibinfo
  {author} {\bibfnamefont {P.}~\bibnamefont {Iacconi}},\ and\ \bibinfo {author}
  {\bibfnamefont {J.~E.}\ \bibnamefont {Butler}},\ }\href
  {https://www.sciencedirect.com/science/article/abs/pii/S0925963506004900}
  {\bibfield  {journal} {\bibinfo  {journal} {Diamond and Related Materials}\
  }\textbf {\bibinfo {volume} {16}},\ \bibinfo {pages} {1062} (\bibinfo {year}
  {2007})}\BibitemShut {NoStop}%
\bibitem [{\citenamefont {Halperin}\ and\ \citenamefont
  {Chen}(1966)}]{halperin1966thermoluminescence}%
  \BibitemOpen
  \bibfield  {author} {\bibinfo {author} {\bibfnamefont {A.}~\bibnamefont
  {Halperin}}\ and\ \bibinfo {author} {\bibfnamefont {R.}~\bibnamefont
  {Chen}},\ }\href
  {https://journals.aps.org/pr/abstract/10.1103/PhysRev.148.839} {\bibfield
  {journal} {\bibinfo  {journal} {Phys. Rev.}\ }\textbf {\bibinfo {volume}
  {148}},\ \bibinfo {pages} {839} (\bibinfo {year} {1966})}\BibitemShut
  {NoStop}%
\bibitem [{\citenamefont {Levinson}\ \emph {et~al.}(1973)\citenamefont
  {Levinson}, \citenamefont {Halperin},\ and\ \citenamefont
  {Bar}}]{levinson1973electrically}%
  \BibitemOpen
  \bibfield  {author} {\bibinfo {author} {\bibfnamefont {J.}~\bibnamefont
  {Levinson}}, \bibinfo {author} {\bibfnamefont {A.}~\bibnamefont {Halperin}},\
  and\ \bibinfo {author} {\bibfnamefont {V.}~\bibnamefont {Bar}},\ }\href
  {https://www.sciencedirect.com/science/article/abs/pii/0022231373900896}
  {\bibfield  {journal} {\bibinfo  {journal} {J. Lumin}\ }\textbf {\bibinfo
  {volume} {6}},\ \bibinfo {pages} {1} (\bibinfo {year} {1973})}\BibitemShut
  {NoStop}%
\bibitem [{\citenamefont {Bourgoin}\ \emph {et~al.}(1978)\citenamefont
  {Bourgoin}, \citenamefont {Massarani},\ and\ \citenamefont
  {Visocekas}}]{bourgoin1978thermally}%
  \BibitemOpen
  \bibfield  {author} {\bibinfo {author} {\bibfnamefont {J.}~\bibnamefont
  {Bourgoin}}, \bibinfo {author} {\bibfnamefont {B.}~\bibnamefont
  {Massarani}},\ and\ \bibinfo {author} {\bibfnamefont {R.}~\bibnamefont
  {Visocekas}},\ }\href
  {https://journals.aps.org/prb/abstract/10.1103/PhysRevB.18.786} {\bibfield
  {journal} {\bibinfo  {journal} {Phys. Rev. B}\ }\textbf {\bibinfo {volume}
  {18}},\ \bibinfo {pages} {786} (\bibinfo {year} {1978})}\BibitemShut
  {NoStop}%
\bibitem [{\citenamefont {Bull}\ and\ \citenamefont
  {Garlick}(1950)}]{bull1950luminescence}%
  \BibitemOpen
  \bibfield  {author} {\bibinfo {author} {\bibfnamefont {C.}~\bibnamefont
  {Bull}}\ and\ \bibinfo {author} {\bibfnamefont {G.~F.~J.}\ \bibnamefont
  {Garlick}},\ }\href
  {https://iopscience.iop.org/article/10.1088/0370-1298/63/11/312/meta}
  {\bibfield  {journal} {\bibinfo  {journal} {Proc. Phys. Soc. A}\ }\textbf
  {\bibinfo {volume} {63}},\ \bibinfo {pages} {1283} (\bibinfo {year}
  {1950})}\BibitemShut {NoStop}%
\bibitem [{\citenamefont {Halperin}\ and\ \citenamefont
  {Nahum}(1961)}]{halperin1961some}%
  \BibitemOpen
  \bibfield  {author} {\bibinfo {author} {\bibfnamefont {A.}~\bibnamefont
  {Halperin}}\ and\ \bibinfo {author} {\bibfnamefont {J.}~\bibnamefont
  {Nahum}},\ }\href
  {https://www.sciencedirect.com/science/article/abs/pii/0022369761901214}
  {\bibfield  {journal} {\bibinfo  {journal} {J. Phys. Chem. Solids}\ }\textbf
  {\bibinfo {volume} {18}},\ \bibinfo {pages} {297} (\bibinfo {year}
  {1961})}\BibitemShut {NoStop}%
\bibitem [{\citenamefont {Shao}\ \emph {et~al.}(2020)\citenamefont {Shao},
  \citenamefont {Lyu}, \citenamefont {Guo}, \citenamefont {Zhang},
  \citenamefont {Zhang}, \citenamefont {Hu},\ and\ \citenamefont
  {Shen}}]{shao2020role}%
  \BibitemOpen
  \bibfield  {author} {\bibinfo {author} {\bibfnamefont {T.}~\bibnamefont
  {Shao}}, \bibinfo {author} {\bibfnamefont {F.}~\bibnamefont {Lyu}}, \bibinfo
  {author} {\bibfnamefont {X.}~\bibnamefont {Guo}}, \bibinfo {author}
  {\bibfnamefont {J.}~\bibnamefont {Zhang}}, \bibinfo {author} {\bibfnamefont
  {H.}~\bibnamefont {Zhang}}, \bibinfo {author} {\bibfnamefont
  {X.}~\bibnamefont {Hu}},\ and\ \bibinfo {author} {\bibfnamefont {A.~H.}\
  \bibnamefont {Shen}},\ }\href
  {https://www.sciencedirect.com/science/article/abs/pii/S0008622320305029}
  {\bibfield  {journal} {\bibinfo  {journal} {Carbon}\ }\textbf {\bibinfo
  {volume} {167}},\ \bibinfo {pages} {888} (\bibinfo {year}
  {2020})}\BibitemShut {NoStop}%
\bibitem [{\citenamefont {Benabdesselam}\ \emph {et~al.}(2000)\citenamefont
  {Benabdesselam}, \citenamefont {Iacconi}, \citenamefont {Briand},
  \citenamefont {Lapraz}, \citenamefont {Gheeraert},\ and\ \citenamefont
  {Deneuville}}]{benabdesselam2000characterisation}%
  \BibitemOpen
  \bibfield  {author} {\bibinfo {author} {\bibfnamefont {M.}~\bibnamefont
  {Benabdesselam}}, \bibinfo {author} {\bibfnamefont {P.}~\bibnamefont
  {Iacconi}}, \bibinfo {author} {\bibfnamefont {D.}~\bibnamefont {Briand}},
  \bibinfo {author} {\bibfnamefont {D.}~\bibnamefont {Lapraz}}, \bibinfo
  {author} {\bibfnamefont {E.}~\bibnamefont {Gheeraert}},\ and\ \bibinfo
  {author} {\bibfnamefont {A.}~\bibnamefont {Deneuville}},\ }\href
  {https://www.sciencedirect.com/science/article/abs/pii/S0925963599002435}
  {\bibfield  {journal} {\bibinfo  {journal} {Diamond and Related Materials}\
  }\textbf {\bibinfo {volume} {9}},\ \bibinfo {pages} {56} (\bibinfo {year}
  {2000})}\BibitemShut {NoStop}%
\bibitem [{\citenamefont {Nahum}\ and\ \citenamefont
  {Halperin}(1963)}]{nahum1963thermoluminescence}%
  \BibitemOpen
  \bibfield  {author} {\bibinfo {author} {\bibfnamefont {J.}~\bibnamefont
  {Nahum}}\ and\ \bibinfo {author} {\bibfnamefont {A.}~\bibnamefont
  {Halperin}},\ }\href
  {https://www.sciencedirect.com/science/article/abs/pii/0022369763900593}
  {\bibfield  {journal} {\bibinfo  {journal} {J. Phys. Chem. Solids}\ }\textbf
  {\bibinfo {volume} {24}},\ \bibinfo {pages} {823} (\bibinfo {year}
  {1963})}\BibitemShut {NoStop}%
\bibitem [{\citenamefont {Howell}\ \emph {et~al.}(2019)\citenamefont {Howell},
  \citenamefont {Collins}, \citenamefont {Loudin}, \citenamefont {Diggle},
  \citenamefont {D'Haenens-Johansson}, \citenamefont {Smit}, \citenamefont
  {Katrusha}, \citenamefont {Butler},\ and\ \citenamefont
  {Nestola}}]{howell2019automated}%
  \BibitemOpen
  \bibfield  {author} {\bibinfo {author} {\bibfnamefont {D.}~\bibnamefont
  {Howell}}, \bibinfo {author} {\bibfnamefont {A.~T.}\ \bibnamefont {Collins}},
  \bibinfo {author} {\bibfnamefont {L.~C.}\ \bibnamefont {Loudin}}, \bibinfo
  {author} {\bibfnamefont {P.~L.}\ \bibnamefont {Diggle}}, \bibinfo {author}
  {\bibfnamefont {U.~F.~S.}\ \bibnamefont {D'Haenens-Johansson}}, \bibinfo
  {author} {\bibfnamefont {K.~V.}\ \bibnamefont {Smit}}, \bibinfo {author}
  {\bibfnamefont {A.~N.}\ \bibnamefont {Katrusha}}, \bibinfo {author}
  {\bibfnamefont {J.~E.}\ \bibnamefont {Butler}},\ and\ \bibinfo {author}
  {\bibfnamefont {F.}~\bibnamefont {Nestola}},\ }\href
  {https://www.sciencedirect.com/science/article/abs/pii/S0925963519303279}
  {\bibfield  {journal} {\bibinfo  {journal} {Diamond and Related Materials}\
  }\textbf {\bibinfo {volume} {96}},\ \bibinfo {pages} {207} (\bibinfo {year}
  {2019})}\BibitemShut {NoStop}%
\bibitem [{\citenamefont {Blank}\ \emph {et~al.}(2007)\citenamefont {Blank},
  \citenamefont {Kuznetsov}, \citenamefont {Nosukhin}, \citenamefont
  {Terentiev},\ and\ \citenamefont {Denisov}}]{blank2007influence}%
  \BibitemOpen
  \bibfield  {author} {\bibinfo {author} {\bibfnamefont {V.~D.}\ \bibnamefont
  {Blank}}, \bibinfo {author} {\bibfnamefont {M.~S.}\ \bibnamefont
  {Kuznetsov}}, \bibinfo {author} {\bibfnamefont {S.~A.}\ \bibnamefont
  {Nosukhin}}, \bibinfo {author} {\bibfnamefont {S.~A.}\ \bibnamefont
  {Terentiev}},\ and\ \bibinfo {author} {\bibfnamefont {V.~N.}\ \bibnamefont
  {Denisov}},\ }\href
  {https://www.sciencedirect.com/science/article/abs/pii/S0925963506005061}
  {\bibfield  {journal} {\bibinfo  {journal} {Diamond and Related Materials}\
  }\textbf {\bibinfo {volume} {16}},\ \bibinfo {pages} {800} (\bibinfo {year}
  {2007})}\BibitemShut {NoStop}%
\bibitem [{\citenamefont {Davies}(1994)}]{davies1994properties}%
  \BibitemOpen
  \bibfield  {author} {\bibinfo {author} {\bibfnamefont {G.}~\bibnamefont
  {Davies}},\ }\href@noop {} {\emph {\bibinfo {title} {Properties and Growth of
  Diamond}}},\ \bibinfo {number} {9}\ (\bibinfo  {publisher} {Inst of
  Engineering \& Technology},\ \bibinfo {year} {1994})\BibitemShut {NoStop}%
\bibitem [{\citenamefont {Burns}\ \emph {et~al.}(1990)\citenamefont {Burns},
  \citenamefont {Cvetkovic}, \citenamefont {Dodge}, \citenamefont {Evans},
  \citenamefont {Rooney}, \citenamefont {Spear},\ and\ \citenamefont
  {Welbourn}}]{burns1990growth}%
  \BibitemOpen
  \bibfield  {author} {\bibinfo {author} {\bibfnamefont {R.~C.}\ \bibnamefont
  {Burns}}, \bibinfo {author} {\bibfnamefont {V.}~\bibnamefont {Cvetkovic}},
  \bibinfo {author} {\bibfnamefont {C.~N.}\ \bibnamefont {Dodge}}, \bibinfo
  {author} {\bibfnamefont {D.~J.~F.}\ \bibnamefont {Evans}}, \bibinfo {author}
  {\bibfnamefont {M.-L.~T.}\ \bibnamefont {Rooney}}, \bibinfo {author}
  {\bibfnamefont {P.~M.}\ \bibnamefont {Spear}},\ and\ \bibinfo {author}
  {\bibfnamefont {C.~M.}\ \bibnamefont {Welbourn}},\ }\href
  {https://www.sciencedirect.com/science/article/pii/0022024890901266}
  {\bibfield  {journal} {\bibinfo  {journal} {J. Cryst. Growth}\ }\textbf
  {\bibinfo {volume} {104}},\ \bibinfo {pages} {257} (\bibinfo {year}
  {1990})}\BibitemShut {NoStop}%
\bibitem [{\citenamefont {Ashfold}\ \emph {et~al.}(2020)\citenamefont
  {Ashfold}, \citenamefont {Goss}, \citenamefont {Green}, \citenamefont {May},
  \citenamefont {Newton},\ and\ \citenamefont {Peaker}}]{ashfold2020nitrogen}%
  \BibitemOpen
  \bibfield  {author} {\bibinfo {author} {\bibfnamefont {M.~N.~R.}\
  \bibnamefont {Ashfold}}, \bibinfo {author} {\bibfnamefont {J.~P.}\
  \bibnamefont {Goss}}, \bibinfo {author} {\bibfnamefont {B.~L.}\ \bibnamefont
  {Green}}, \bibinfo {author} {\bibfnamefont {P.~W.}\ \bibnamefont {May}},
  \bibinfo {author} {\bibfnamefont {M.~E.}\ \bibnamefont {Newton}},\ and\
  \bibinfo {author} {\bibfnamefont {C.~V.}\ \bibnamefont {Peaker}},\ }\href
  {https://pubs.acs.org/doi/abs/10.1021/acs.chemrev.9b00518} {\bibfield
  {journal} {\bibinfo  {journal} {Chem. Rev.}\ }\textbf {\bibinfo {volume}
  {120}},\ \bibinfo {pages} {5745} (\bibinfo {year} {2020})}\BibitemShut
  {NoStop}%
\bibitem [{\citenamefont {Collins}(2002)}]{collins2002fermi}%
  \BibitemOpen
  \bibfield  {author} {\bibinfo {author} {\bibfnamefont {A.~T.}\ \bibnamefont
  {Collins}},\ }\href
  {https://iopscience.iop.org/article/10.1088/0953-8984/14/14/307/meta}
  {\bibfield  {journal} {\bibinfo  {journal} {J. Phys. Condens.}\ }\textbf
  {\bibinfo {volume} {14}},\ \bibinfo {pages} {3743} (\bibinfo {year}
  {2002})}\BibitemShut {NoStop}%
\bibitem [{\citenamefont {Jones}\ \emph {et~al.}(2009)\citenamefont {Jones},
  \citenamefont {Goss},\ and\ \citenamefont {Briddon}}]{jones2009acceptor}%
  \BibitemOpen
  \bibfield  {author} {\bibinfo {author} {\bibfnamefont {R.}~\bibnamefont
  {Jones}}, \bibinfo {author} {\bibfnamefont {J.~P.}\ \bibnamefont {Goss}},\
  and\ \bibinfo {author} {\bibfnamefont {P.~R.}\ \bibnamefont {Briddon}},\
  }\href {https://journals.aps.org/prb/abstract/10.1103/PhysRevB.80.033205}
  {\bibfield  {journal} {\bibinfo  {journal} {Phys. Rev. B}\ }\textbf {\bibinfo
  {volume} {80}},\ \bibinfo {pages} {033205} (\bibinfo {year}
  {2009})}\BibitemShut {NoStop}%
\bibitem [{\citenamefont {Ulbricht}\ \emph {et~al.}(2011)\citenamefont
  {Ulbricht}, \citenamefont {Van Der~Post}, \citenamefont {Goss}, \citenamefont
  {Briddon}, \citenamefont {Jones}, \citenamefont {Khan},\ and\ \citenamefont
  {Bonn}}]{ulbricht2011single}%
  \BibitemOpen
  \bibfield  {author} {\bibinfo {author} {\bibfnamefont {R.}~\bibnamefont
  {Ulbricht}}, \bibinfo {author} {\bibfnamefont {S.~T.}\ \bibnamefont {Van
  Der~Post}}, \bibinfo {author} {\bibfnamefont {J.~P.}\ \bibnamefont {Goss}},
  \bibinfo {author} {\bibfnamefont {P.~R.}\ \bibnamefont {Briddon}}, \bibinfo
  {author} {\bibfnamefont {R.}~\bibnamefont {Jones}}, \bibinfo {author}
  {\bibfnamefont {R.~U.~A.}\ \bibnamefont {Khan}},\ and\ \bibinfo {author}
  {\bibfnamefont {M.}~\bibnamefont {Bonn}},\ }\href
  {https://journals.aps.org/prb/abstract/10.1103/PhysRevB.84.165202} {\bibfield
   {journal} {\bibinfo  {journal} {Phys. Rev. B}\ }\textbf {\bibinfo {volume}
  {84}},\ \bibinfo {pages} {165202} (\bibinfo {year} {2011})}\BibitemShut
  {NoStop}%
\bibitem [{\citenamefont {Collins}(2010)}]{collins2010}%
  \BibitemOpen
  \bibfield  {author} {\bibinfo {author} {\bibfnamefont {A.~T.}\ \bibnamefont
  {Collins}},\ }in\ \href@noop {} {\emph {\bibinfo {booktitle} {Diamond
  Conference, University of Warwick, July}}}\ (\bibinfo {year} {2010})\ pp.\
  \bibinfo {pages} {13--16}\BibitemShut {NoStop}%
\bibitem [{\citenamefont {Breeze}(2016)}]{breeze2016electron}%
  \BibitemOpen
  \bibfield  {author} {\bibinfo {author} {\bibfnamefont {B.~G.}\ \bibnamefont
  {Breeze}},\ }\emph {\bibinfo {title} {Electron paramagnetic resonance studies
  of point defects in diamond: quantification, spin polarisation and
  preferential orientation}},\ \href
  {https://ethos.bl.uk/OrderDetails.do?uin=uk.bl.ethos.720455} {Ph.D. thesis},\
  \bibinfo  {school} {University of Warwick} (\bibinfo {year}
  {2016})\BibitemShut {NoStop}%
\bibitem [{\citenamefont {Li}\ \emph {et~al.}(2016)\citenamefont {Li},
  \citenamefont {Fan}, \citenamefont {Chen},\ and\ \citenamefont
  {Li}}]{li2016diamond}%
  \BibitemOpen
  \bibfield  {author} {\bibinfo {author} {\bibfnamefont {J.}~\bibnamefont
  {Li}}, \bibinfo {author} {\bibfnamefont {C.}~\bibnamefont {Fan}}, \bibinfo
  {author} {\bibfnamefont {S.}~\bibnamefont {Chen}},\ and\ \bibinfo {author}
  {\bibfnamefont {G.}~\bibnamefont {Li}},\ }\href
  {http://www.sdim.cn/ngdtc/uploads/soft/200317/1-20031G62350.pdf} {\bibfield
  {journal} {\bibinfo  {journal} {The Journal of Gemmology}\ }\textbf {\bibinfo
  {volume} {35}},\ \bibinfo {pages} {248} (\bibinfo {year} {2016})}\BibitemShut
  {NoStop}%
\bibitem [{\citenamefont {Eaton-Maga{\~n}a}(2016)}]{eaton2016decay}%
  \BibitemOpen
  \bibfield  {author} {\bibinfo {author} {\bibfnamefont {S.}~\bibnamefont
  {Eaton-Maga{\~n}a}},\ }\href
  {https://www.researchgate.net/profile/Sally-Eaton-Magana/publication/313890582_Decay_Kinetics_of_Boron-Related_Peak_in_IR_Absorption_of_Natural_Diamond/links/5ac4d8340f7e9becc9d620e8/Decay-Kinetics-of-Boron-Related-Peak-in-IR-Absorption-of-Natural-Diamond.pdf}
  {\bibfield  {journal} {\bibinfo  {journal} {GEMS \& GEMOLOGY}\ }\textbf
  {\bibinfo {volume} {52}},\ \bibinfo {pages} {412} (\bibinfo {year}
  {2016})}\BibitemShut {NoStop}%
\bibitem [{\citenamefont {Dobrowolska}\ \emph {et~al.}(2014)\citenamefont
  {Dobrowolska}, \citenamefont {Bos},\ and\ \citenamefont
  {Dorenbos}}]{dobrowolska2014electron}%
  \BibitemOpen
  \bibfield  {author} {\bibinfo {author} {\bibfnamefont {A.}~\bibnamefont
  {Dobrowolska}}, \bibinfo {author} {\bibfnamefont {A.~J.~J.}\ \bibnamefont
  {Bos}},\ and\ \bibinfo {author} {\bibfnamefont {P.}~\bibnamefont
  {Dorenbos}},\ }\href
  {https://iopscience.iop.org/article/10.1088/0022-3727/47/33/335301/meta}
  {\bibfield  {journal} {\bibinfo  {journal} {J. Phys. D: Appl. Phys.}\
  }\textbf {\bibinfo {volume} {47}},\ \bibinfo {pages} {335301} (\bibinfo
  {year} {2014})}\BibitemShut {NoStop}%
\bibitem [{\citenamefont {Berberan-Santos}\ \emph
  {et~al.}(2005{\natexlab{a}})\citenamefont {Berberan-Santos}, \citenamefont
  {Bodunov},\ and\ \citenamefont {Valeur}}]{berberan2005mathematical}%
  \BibitemOpen
  \bibfield  {author} {\bibinfo {author} {\bibfnamefont {M.~N.}\ \bibnamefont
  {Berberan-Santos}}, \bibinfo {author} {\bibfnamefont {E.~N.}\ \bibnamefont
  {Bodunov}},\ and\ \bibinfo {author} {\bibfnamefont {B.}~\bibnamefont
  {Valeur}},\ }\href
  {https://www.sciencedirect.com/science/article/abs/pii/S0301010405001175}
  {\bibfield  {journal} {\bibinfo  {journal} {Chem. Phys.}\ }\textbf {\bibinfo
  {volume} {315}},\ \bibinfo {pages} {171} (\bibinfo {year}
  {2005}{\natexlab{a}})}\BibitemShut {NoStop}%
\bibitem [{\citenamefont {Chen}(2003)}]{chen2003apparent}%
  \BibitemOpen
  \bibfield  {author} {\bibinfo {author} {\bibfnamefont {R.}~\bibnamefont
  {Chen}},\ }\href
  {https://www.sciencedirect.com/science/article/abs/pii/S0022231302006014}
  {\bibfield  {journal} {\bibinfo  {journal} {J. Lumin.}\ }\textbf {\bibinfo
  {volume} {102--103}},\ \bibinfo {pages} {510} (\bibinfo {year}
  {2003})}\BibitemShut {NoStop}%
\bibitem [{\citenamefont {Vedda}\ and\ \citenamefont
  {Fasoli}(2018)}]{vedda2018tunneling}%
  \BibitemOpen
  \bibfield  {author} {\bibinfo {author} {\bibfnamefont {A.}~\bibnamefont
  {Vedda}}\ and\ \bibinfo {author} {\bibfnamefont {M.}~\bibnamefont {Fasoli}},\
  }\href
  {https://www.sciencedirect.com/science/article/abs/pii/S1350448718301835}
  {\bibfield  {journal} {\bibinfo  {journal} {Radiat. Meas.}\ }\textbf
  {\bibinfo {volume} {118}},\ \bibinfo {pages} {86} (\bibinfo {year}
  {2018})}\BibitemShut {NoStop}%
\bibitem [{\citenamefont {Medlin}(1961)}]{medlin1961decay}%
  \BibitemOpen
  \bibfield  {author} {\bibinfo {author} {\bibfnamefont {W.~L.}\ \bibnamefont
  {Medlin}},\ }\href
  {https://journals.aps.org/pr/abstract/10.1103/PhysRev.122.837} {\bibfield
  {journal} {\bibinfo  {journal} {Phys. Rev.}\ }\textbf {\bibinfo {volume}
  {122}},\ \bibinfo {pages} {837} (\bibinfo {year} {1961})}\BibitemShut
  {NoStop}%
\bibitem [{\citenamefont {Berberan-Santos}\ \emph
  {et~al.}(2005{\natexlab{b}})\citenamefont {Berberan-Santos}, \citenamefont
  {Bodunov},\ and\ \citenamefont {Valeur}}]{berberan2005mathematical2}%
  \BibitemOpen
  \bibfield  {author} {\bibinfo {author} {\bibfnamefont {M.~N.}\ \bibnamefont
  {Berberan-Santos}}, \bibinfo {author} {\bibfnamefont {E.~N.}\ \bibnamefont
  {Bodunov}},\ and\ \bibinfo {author} {\bibfnamefont {B.}~\bibnamefont
  {Valeur}},\ }\href
  {https://www.sciencedirect.com/science/article/abs/pii/S0301010405002211}
  {\bibfield  {journal} {\bibinfo  {journal} {Chem. Phys.}\ }\textbf {\bibinfo
  {volume} {317}},\ \bibinfo {pages} {57} (\bibinfo {year}
  {2005}{\natexlab{b}})}\BibitemShut {NoStop}%
\bibitem [{\citenamefont {Hopfield}\ \emph {et~al.}(1963)\citenamefont
  {Hopfield}, \citenamefont {Thomas},\ and\ \citenamefont
  {Gershenzon}}]{hopfield1963pair}%
  \BibitemOpen
  \bibfield  {author} {\bibinfo {author} {\bibfnamefont {J.~J.}\ \bibnamefont
  {Hopfield}}, \bibinfo {author} {\bibfnamefont {D.~G.}\ \bibnamefont
  {Thomas}},\ and\ \bibinfo {author} {\bibfnamefont {M.}~\bibnamefont
  {Gershenzon}},\ }\href
  {https://journals.aps.org/prl/abstract/10.1103/PhysRevLett.10.162} {\bibfield
   {journal} {\bibinfo  {journal} {Phys. Rev. Lett.}\ }\textbf {\bibinfo
  {volume} {10}},\ \bibinfo {pages} {162} (\bibinfo {year} {1963})}\BibitemShut
  {NoStop}%
\bibitem [{\citenamefont {Sun}\ \emph {et~al.}(2011)\citenamefont {Sun},
  \citenamefont {Ivanov}, \citenamefont {Juillaguet},\ and\ \citenamefont
  {Camassel}}]{sun2011splitting}%
  \BibitemOpen
  \bibfield  {author} {\bibinfo {author} {\bibfnamefont {J.~W.}\ \bibnamefont
  {Sun}}, \bibinfo {author} {\bibfnamefont {I.~G.}\ \bibnamefont {Ivanov}},
  \bibinfo {author} {\bibfnamefont {S.}~\bibnamefont {Juillaguet}},\ and\
  \bibinfo {author} {\bibfnamefont {J.}~\bibnamefont {Camassel}},\ }\href
  {https://journals.aps.org/prb/abstract/10.1103/PhysRevB.83.195201} {\bibfield
   {journal} {\bibinfo  {journal} {Phys. Rev. B}\ }\textbf {\bibinfo {volume}
  {83}},\ \bibinfo {pages} {195201} (\bibinfo {year} {2011})}\BibitemShut
  {NoStop}%
\bibitem [{\citenamefont {Dischler}\ \emph {et~al.}(1994)\citenamefont
  {Dischler}, \citenamefont {Rothemund}, \citenamefont {Wild}, \citenamefont
  {Locher}, \citenamefont {Biebl},\ and\ \citenamefont {Koidl}}]{dischler1994}%
  \BibitemOpen
  \bibfield  {author} {\bibinfo {author} {\bibfnamefont {B.}~\bibnamefont
  {Dischler}}, \bibinfo {author} {\bibfnamefont {W.}~\bibnamefont {Rothemund}},
  \bibinfo {author} {\bibfnamefont {C.}~\bibnamefont {Wild}}, \bibinfo {author}
  {\bibfnamefont {R.}~\bibnamefont {Locher}}, \bibinfo {author} {\bibfnamefont
  {H.}~\bibnamefont {Biebl}},\ and\ \bibinfo {author} {\bibfnamefont
  {P.}~\bibnamefont {Koidl}},\ }\href
  {https://journals.aps.org/prb/abstract/10.1103/PhysRevB.49.1685} {\bibfield
  {journal} {\bibinfo  {journal} {Phys. Rev. B}\ }\textbf {\bibinfo {volume}
  {49}},\ \bibinfo {pages} {1685} (\bibinfo {year} {1994})}\BibitemShut
  {NoStop}%
\bibitem [{\citenamefont {{\v{S}}{\v{c}}ajev}\ \emph
  {et~al.}(2013)\citenamefont {{\v{S}}{\v{c}}ajev}, \citenamefont {Trinkler},
  \citenamefont {Berzina}, \citenamefont {Ivakin},\ and\ \citenamefont
  {Jara{\v{s}}i{\=u}nas}}]{svcajev2013}%
  \BibitemOpen
  \bibfield  {author} {\bibinfo {author} {\bibfnamefont {P.}~\bibnamefont
  {{\v{S}}{\v{c}}ajev}}, \bibinfo {author} {\bibfnamefont {L.}~\bibnamefont
  {Trinkler}}, \bibinfo {author} {\bibfnamefont {B.}~\bibnamefont {Berzina}},
  \bibinfo {author} {\bibfnamefont {E.}~\bibnamefont {Ivakin}},\ and\ \bibinfo
  {author} {\bibfnamefont {K.}~\bibnamefont {Jara{\v{s}}i{\=u}nas}},\ }\href
  {https://www.sciencedirect.com/science/article/abs/pii/S0925963513000678}
  {\bibfield  {journal} {\bibinfo  {journal} {Diamond and Related Materials}\
  }\textbf {\bibinfo {volume} {36}},\ \bibinfo {pages} {35} (\bibinfo {year}
  {2013})}\BibitemShut {NoStop}%
\bibitem [{\citenamefont {Cox}\ \emph {et~al.}(1994)\citenamefont {Cox},
  \citenamefont {Newton},\ and\ \citenamefont {Baker}}]{cox1994}%
  \BibitemOpen
  \bibfield  {author} {\bibinfo {author} {\bibfnamefont {A.}~\bibnamefont
  {Cox}}, \bibinfo {author} {\bibfnamefont {M.~E.}\ \bibnamefont {Newton}},\
  and\ \bibinfo {author} {\bibfnamefont {J.~M.}\ \bibnamefont {Baker}},\ }\href
  {https://iopscience.iop.org/article/10.1088/0953-8984/6/2/025/meta}
  {\bibfield  {journal} {\bibinfo  {journal} {J. Phys.: Condens. Matter}\
  }\textbf {\bibinfo {volume} {6}},\ \bibinfo {pages} {551} (\bibinfo {year}
  {1994})}\BibitemShut {NoStop}%
\bibitem [{\citenamefont {Goss}\ and\ \citenamefont
  {Briddon}(2006)}]{goss2006}%
  \BibitemOpen
  \bibfield  {author} {\bibinfo {author} {\bibfnamefont {J.~P.}\ \bibnamefont
  {Goss}}\ and\ \bibinfo {author} {\bibfnamefont {P.~R.}\ \bibnamefont
  {Briddon}},\ }\href
  {https://journals.aps.org/prb/abstract/10.1103/PhysRevB.73.085204} {\bibfield
   {journal} {\bibinfo  {journal} {Phys. Rev. B}\ }\textbf {\bibinfo {volume}
  {73}},\ \bibinfo {pages} {085204} (\bibinfo {year} {2006})}\BibitemShut
  {NoStop}%
\bibitem [{\citenamefont {Mortet}\ \emph {et~al.}(2018)\citenamefont {Mortet},
  \citenamefont {Taylor}, \citenamefont {{\v{Z}}ivcov{\'a}}, \citenamefont
  {Machon}, \citenamefont {Frank}, \citenamefont {Hub{\'\i}k}, \citenamefont
  {Tr{\'e}mouilles},\ and\ \citenamefont {Kavan}}]{mortet2018}%
  \BibitemOpen
  \bibfield  {author} {\bibinfo {author} {\bibfnamefont {V.}~\bibnamefont
  {Mortet}}, \bibinfo {author} {\bibfnamefont {A.}~\bibnamefont {Taylor}},
  \bibinfo {author} {\bibfnamefont {Z.~V.}\ \bibnamefont {{\v{Z}}ivcov{\'a}}},
  \bibinfo {author} {\bibfnamefont {D.}~\bibnamefont {Machon}}, \bibinfo
  {author} {\bibfnamefont {O.}~\bibnamefont {Frank}}, \bibinfo {author}
  {\bibfnamefont {P.}~\bibnamefont {Hub{\'\i}k}}, \bibinfo {author}
  {\bibfnamefont {D.}~\bibnamefont {Tr{\'e}mouilles}},\ and\ \bibinfo {author}
  {\bibfnamefont {L.}~\bibnamefont {Kavan}},\ }\href
  {https://www.sciencedirect.com/science/article/abs/pii/S0925963518303467}
  {\bibfield  {journal} {\bibinfo  {journal} {Diamond and Related Materials}\
  }\textbf {\bibinfo {volume} {88}},\ \bibinfo {pages} {163} (\bibinfo {year}
  {2018})}\BibitemShut {NoStop}%
\bibitem [{\citenamefont {Alkauskas}\ \emph {et~al.}(2016)\citenamefont
  {Alkauskas}, \citenamefont {McCluskey},\ and\ \citenamefont {Van~de
  Walle}}]{alkauskas2016}%
  \BibitemOpen
  \bibfield  {author} {\bibinfo {author} {\bibfnamefont {A.}~\bibnamefont
  {Alkauskas}}, \bibinfo {author} {\bibfnamefont {M.~D.}\ \bibnamefont
  {McCluskey}},\ and\ \bibinfo {author} {\bibfnamefont {C.~G.}\ \bibnamefont
  {Van~de Walle}},\ }\href
  {https://aip.scitation.org/doi/abs/10.1063/1.4948245} {\bibfield  {journal}
  {\bibinfo  {journal} {J. Appl. Phys.}\ }\textbf {\bibinfo {volume} {119}},\
  \bibinfo {pages} {181101} (\bibinfo {year} {2016})}\BibitemShut {NoStop}%
\bibitem [{\citenamefont {Franceschetti}\ and\ \citenamefont
  {Pantelides}(2003)}]{franceschetti2003}%
  \BibitemOpen
  \bibfield  {author} {\bibinfo {author} {\bibfnamefont {A.}~\bibnamefont
  {Franceschetti}}\ and\ \bibinfo {author} {\bibfnamefont {S.~T.}\ \bibnamefont
  {Pantelides}},\ }\href
  {https://journals.aps.org/prb/abstract/10.1103/PhysRevB.68.033313} {\bibfield
   {journal} {\bibinfo  {journal} {Phys. Rev. B}\ }\textbf {\bibinfo {volume}
  {68}},\ \bibinfo {pages} {033313} (\bibinfo {year} {2003})}\BibitemShut
  {NoStop}%
\bibitem [{\citenamefont {Ammerlaan}\ and\ \citenamefont
  {Burgemeister}(1981)}]{ammerlaan1981}%
  \BibitemOpen
  \bibfield  {author} {\bibinfo {author} {\bibfnamefont {C.~A.~J.}\
  \bibnamefont {Ammerlaan}}\ and\ \bibinfo {author} {\bibfnamefont {E.~A.}\
  \bibnamefont {Burgemeister}},\ }\href
  {https://journals.aps.org/prl/abstract/10.1103/PhysRevLett.47.954} {\bibfield
   {journal} {\bibinfo  {journal} {Phys. Rev. Lett.}\ }\textbf {\bibinfo
  {volume} {47}},\ \bibinfo {pages} {954} (\bibinfo {year} {1981})}\BibitemShut
  {NoStop}%
\bibitem [{\citenamefont {Alkauskas}\ \emph {et~al.}(2012)\citenamefont
  {Alkauskas}, \citenamefont {Lyons}, \citenamefont {Steiauf},\ and\
  \citenamefont {Van~de Walle}}]{alkauskas2012}%
  \BibitemOpen
  \bibfield  {author} {\bibinfo {author} {\bibfnamefont {A.}~\bibnamefont
  {Alkauskas}}, \bibinfo {author} {\bibfnamefont {J.~L.}\ \bibnamefont
  {Lyons}}, \bibinfo {author} {\bibfnamefont {D.}~\bibnamefont {Steiauf}},\
  and\ \bibinfo {author} {\bibfnamefont {C.~G.}\ \bibnamefont {Van~de Walle}},\
  }\href {https://journals.aps.org/prl/abstract/10.1103/PhysRevLett.109.267401}
  {\bibfield  {journal} {\bibinfo  {journal} {Phys. Rev. Lett.}\ }\textbf
  {\bibinfo {volume} {109}},\ \bibinfo {pages} {267401} (\bibinfo {year}
  {2012})}\BibitemShut {NoStop}%
\bibitem [{\citenamefont {Avouris}\ and\ \citenamefont
  {Morgan}(1981)}]{avouris1981tunneling}%
  \BibitemOpen
  \bibfield  {author} {\bibinfo {author} {\bibfnamefont {P.}~\bibnamefont
  {Avouris}}\ and\ \bibinfo {author} {\bibfnamefont {T.~N.}\ \bibnamefont
  {Morgan}},\ }\href {https://aip.scitation.org/doi/abs/10.1063/1.441677}
  {\bibfield  {journal} {\bibinfo  {journal} {J. Chem. Phys.}\ }\textbf
  {\bibinfo {volume} {74}},\ \bibinfo {pages} {4347} (\bibinfo {year}
  {1981})}\BibitemShut {NoStop}%
\bibitem [{\citenamefont {Avouris}\ \emph {et~al.}(1981)\citenamefont
  {Avouris}, \citenamefont {Chang}, \citenamefont {Dove}, \citenamefont
  {Morgan},\ and\ \citenamefont {Thefaine}}]{avouris1981trapping}%
  \BibitemOpen
  \bibfield  {author} {\bibinfo {author} {\bibfnamefont {P.}~\bibnamefont
  {Avouris}}, \bibinfo {author} {\bibfnamefont {I.~F.}\ \bibnamefont {Chang}},
  \bibinfo {author} {\bibfnamefont {D.}~\bibnamefont {Dove}}, \bibinfo {author}
  {\bibfnamefont {T.~N.}\ \bibnamefont {Morgan}},\ and\ \bibinfo {author}
  {\bibfnamefont {Y.}~\bibnamefont {Thefaine}},\ }\href
  {https://link.springer.com/article/10.1007/BF02661006} {\bibfield  {journal}
  {\bibinfo  {journal} {J. Electron. Mater.}\ }\textbf {\bibinfo {volume}
  {10}},\ \bibinfo {pages} {887} (\bibinfo {year} {1981})}\BibitemShut
  {NoStop}%
\bibitem [{\citenamefont {Phillips}(1996)}]{phillips1996}%
  \BibitemOpen
  \bibfield  {author} {\bibinfo {author} {\bibfnamefont {J.~C.}\ \bibnamefont
  {Phillips}},\ }\href
  {https://iopscience.iop.org/article/10.1088/0034-4885/59/9/003} {\bibfield
  {journal} {\bibinfo  {journal} {Rep. Prog. Phys.}\ }\textbf {\bibinfo
  {volume} {59}},\ \bibinfo {pages} {1133} (\bibinfo {year}
  {1996})}\BibitemShut {NoStop}%
\bibitem [{\citenamefont {Collins}\ and\ \citenamefont
  {Williams}(1971)}]{collins1971nature}%
  \BibitemOpen
  \bibfield  {author} {\bibinfo {author} {\bibfnamefont {A.~T.}\ \bibnamefont
  {Collins}}\ and\ \bibinfo {author} {\bibfnamefont {A.~W.~S.}\ \bibnamefont
  {Williams}},\ }\href
  {https://iopscience.iop.org/article/10.1088/0022-3719/4/13/030/meta}
  {\bibfield  {journal} {\bibinfo  {journal} {J. Phys. C: Solid State Phys.}\
  }\textbf {\bibinfo {volume} {4}},\ \bibinfo {pages} {1789} (\bibinfo {year}
  {1971})}\BibitemShut {NoStop}%
\bibitem [{\citenamefont {Chrenko}(1973)}]{chrenko1973boron}%
  \BibitemOpen
  \bibfield  {author} {\bibinfo {author} {\bibfnamefont {R.~M.}\ \bibnamefont
  {Chrenko}},\ }\href
  {https://journals.aps.org/prb/abstract/10.1103/PhysRevB.7.4560} {\bibfield
  {journal} {\bibinfo  {journal} {Phys. Rev. B}\ }\textbf {\bibinfo {volume}
  {7}},\ \bibinfo {pages} {4560} (\bibinfo {year} {1973})}\BibitemShut
  {NoStop}%
\bibitem [{\citenamefont {Collins}(1999)}]{collins1999things}%
  \BibitemOpen
  \bibfield  {author} {\bibinfo {author} {\bibfnamefont {A.~T.}\ \bibnamefont
  {Collins}},\ }\href
  {https://www.sciencedirect.com/science/article/abs/pii/S0925963599000138}
  {\bibfield  {journal} {\bibinfo  {journal} {Diamond and Related Materials}\
  }\textbf {\bibinfo {volume} {8}},\ \bibinfo {pages} {1455} (\bibinfo {year}
  {1999})}\BibitemShut {NoStop}%
\bibitem [{\citenamefont {Lawson}\ \emph {et~al.}(1998)\citenamefont {Lawson},
  \citenamefont {Fisher}, \citenamefont {Hunt},\ and\ \citenamefont
  {Newton}}]{lawson1998existence}%
  \BibitemOpen
  \bibfield  {author} {\bibinfo {author} {\bibfnamefont {S.~C.}\ \bibnamefont
  {Lawson}}, \bibinfo {author} {\bibfnamefont {D.}~\bibnamefont {Fisher}},
  \bibinfo {author} {\bibfnamefont {D.~C.}\ \bibnamefont {Hunt}},\ and\
  \bibinfo {author} {\bibfnamefont {M.~E.}\ \bibnamefont {Newton}},\ }\href
  {https://iopscience.iop.org/article/10.1088/0953-8984/10/27/016/meta}
  {\bibfield  {journal} {\bibinfo  {journal} {J. Phys.: Condens. Matter}\
  }\textbf {\bibinfo {volume} {10}},\ \bibinfo {pages} {6171} (\bibinfo {year}
  {1998})}\BibitemShut {NoStop}%
\bibitem [{\citenamefont {Huang}\ and\ \citenamefont {Rhys}(1950)}]{huang1950}%
  \BibitemOpen
  \bibfield  {author} {\bibinfo {author} {\bibfnamefont {K.}~\bibnamefont
  {Huang}}\ and\ \bibinfo {author} {\bibfnamefont {A.}~\bibnamefont {Rhys}},\
  }\href {https://royalsocietypublishing.org/doi/abs/10.1098/rspa.1950.0184}
  {\bibfield  {journal} {\bibinfo  {journal} {Proc. R. Soc. A}\ }\textbf
  {\bibinfo {volume} {204}},\ \bibinfo {pages} {406} (\bibinfo {year}
  {1950})}\BibitemShut {NoStop}%
\bibitem [{\citenamefont {Van~Zeghbroeck}(2011)}]{van2011principles}%
  \BibitemOpen
  \bibfield  {author} {\bibinfo {author} {\bibfnamefont {B.~J.}\ \bibnamefont
  {Van~Zeghbroeck}},\ }\href@noop {} {\emph {\bibinfo {title} {Principles of
  semiconductor devices}}}\ (\bibinfo  {publisher} {University of Colorado},\
  \bibinfo {year} {2011})\BibitemShut {NoStop}%
\end{thebibliography}%

\end{document}